\documentclass[%
aip,
 amsmath,amssymb,
 preprint,%
]{revtex4-2}

\usepackage{graphicx}
\usepackage{dcolumn}
\usepackage{bm}

\usepackage[utf8]{inputenc}
\usepackage[T1]{fontenc}
\usepackage{mathptmx}
\usepackage{hyperref} 
\usepackage{xcolor}

\begin{document}


\title[]{Ultrafast spin-currents and charge conversion at 3\textit{d}-5\textit{d} interfaces probed by time-domain terahertz spectroscopy}

\author{T.~H. Dang}
\affiliation{Unité Mixte de Physique CNRS/Thales, Université Paris-Saclay, 91767 Palaiseau, France}
\author{J. Hawecker}
\affiliation{Laboratoire de Physique de l’Ecole Normale Supérieure, ENS, Université PSL, CNRS, Sorbonne Université, Université de Paris, Paris, France}
\author{E. Rongione}
\affiliation{Unité Mixte de Physique CNRS/Thales, Université Paris-Saclay, 91767 Palaiseau, France}
\author{G. Baez Flores}
\affiliation{Department of Physics and Astronomy and Nebraska Center for Materials and Nanoscience, University of Nebraska-Lincoln, Lincoln, Nebraska 68588, USA}
\author{D.~Q. To}
\affiliation{Unité Mixte de Physique CNRS/Thales, Université Paris-Saclay, 91767 Palaiseau, France}
\author{J.~C. Rojas-Sanchez}
\affiliation{Unité Mixte de Physique CNRS/Thales, Université Paris-Saclay, 91767 Palaiseau, France}
\author{H. Nong}
\affiliation{Laboratoire de Physique de l’Ecole Normale Supérieure, ENS, Université PSL, CNRS, Sorbonne Université, Université de Paris, Paris, France}
\author{J. Mangeney}
\affiliation{Laboratoire de Physique de l’Ecole Normale Supérieure, ENS, Université PSL, CNRS, Sorbonne Université, Université de Paris, Paris, France}
\author{J. Tignon}
\affiliation{Laboratoire de Physique de l’Ecole Normale Supérieure, ENS, Université PSL, CNRS, Sorbonne Université, Université de Paris, Paris, France}
\author{F. Godel}
\affiliation{Unité Mixte de Physique CNRS/Thales, Université Paris-Saclay, 91767 Palaiseau, France}
\author{S. Collin}
\affiliation{Unité Mixte de Physique CNRS/Thales, Université Paris-Saclay, 91767 Palaiseau, France}
\author{P. Seneor}
\affiliation{Unité Mixte de Physique CNRS/Thales, Université Paris-Saclay, 91767 Palaiseau, France}
\author{M. Bibes}
\affiliation{Unité Mixte de Physique CNRS/Thales, Université Paris-Saclay, 91767 Palaiseau, France}
\author{A. Fert}
\affiliation{Unité Mixte de Physique CNRS/Thales, Université Paris-Saclay, 91767 Palaiseau, France}
\author{M. Anane}
\affiliation{Unité Mixte de Physique CNRS/Thales, Université Paris-Saclay, 91767 Palaiseau, France}
\author{J.-M. George}
\affiliation{Unité Mixte de Physique CNRS/Thales, Université Paris-Saclay, 91767 Palaiseau, France}
\author{L. Vila}
\affiliation{Université Grenoble Alpes, CEA, CNRS, Grenoble INP, Spintec, 38000 Grenoble, France}
\author{M. Cosset-Cheneau}
\affiliation{Université Grenoble Alpes, CEA, CNRS, Grenoble INP, Spintec, 38000 Grenoble, France}
\author{D. Dolfi}
\affiliation{Thales Research \& Technology, 1 Avenue Augustin Fresnel, 91767 Palaiseau, France}
\author{R. Lebrun}
\affiliation{Unité Mixte de Physique CNRS/Thales, Université Paris-Saclay, 91767 Palaiseau, France}
\author{P. Bortolotti}
\affiliation{Unité Mixte de Physique CNRS/Thales, Université Paris-Saclay, 91767 Palaiseau, France}
\author{K. Belashchenko}
\affiliation{Department of Physics and Astronomy and Nebraska Center for Materials and Nanoscience, University of Nebraska-Lincoln, Lincoln, Nebraska 68588, USA}
\author{S. Dhillon}
\affiliation{Laboratoire de Physique de l’Ecole Normale Supérieure, ENS, Université PSL, CNRS, Sorbonne Université, Université de Paris, Paris, France}
\author{H. Jaffrès}
\affiliation{Unité Mixte de Physique CNRS/Thales, Université Paris-Saclay, 91767 Palaiseau, France}

\date{\today}

\begin{abstract}
\textcolor{black}{Spintronic structures are extensively investigated for their spin orbit torque properties, required for magnetic commutation functionalities. Current progress in these materials is dependent on the interface engineering for the optimization of spin transmission. Here, we advance the analysis of ultrafast spin-charge conversion phenomena at ferromagnetic - transition metal interfaces due to their inverse spin-Hall effect properties. In particular the intrinsic inverse spin Hall effect of Pt-based systems and extrinsic inverse spin-Hall effect of Au:W and Au:Ta in NiFe/Au:(W,Ta) bilayers are investigated. The spin-charge conversion is probed by complementary techniques - ultrafast THz time domain spectroscopy in the dynamic regime for THz pulse emission and ferromagnetic resonance spin-pumping measurements in the GHz regime in the steady state - to determine the role played by the material properties, resistivities, spin transmission at metallic interfaces and spin-flip rates. These measurements show the correspondence between the THz time domain spectroscopy and ferromagnetic spin-pumping for the different set of samples in term of the spin mixing conductance. The latter quantity is a critical parameter, determining the strength of the THz emission from spintronic interfaces. This is further supported by ab-initio calculations, simulations and analysis of the spin-diffusion and spin relaxation of carriers within the multilayers in the time domain, permitting to determine the main trends and the role of spin transmission at interfaces. This work illustrates that time domain spectroscopy for spin-based THz emission is a powerful technique to probe spin-dynamics at active spintronic interfaces and to extract key material properties for spin-charge conversion.}
\end{abstract}

\maketitle

\begin{quotation}
\end{quotation}

\section*{\label{Introduction}Introduction\protect}

The terahertz (THz) frequency range of the electromagnetic spectrum is generally defined as extending from 0.3 to 10 THz. It represents a spectral window that offers rich opportunities for advanced industrial applications in fields such as quality control, spectroscopy, imaging, medical diagnostics, security, telecommunications and high-speed electronics. A range of promising THz sources techniques exist, as well new technologies that are being developed~\cite{Dhillon_2017} alongwith non-destructive testing, biodetection, cancer imaging and microscopy using spintronics emitters~\cite{chen2020}. A widespread method is based on THz time domain spectroscopy (TDS) where THz pulses are typically generated from nonlinear optical crystals \textit{via} optical rectification or from photoconductive switches with ultrafast transient currents when excited by an ultrashort near-infrared femtosecond oscillator~\cite{Hilton04}. Nonetheless, the polar properties of the crystals and  semiconductors used can strongly attenuate the THz radiation around optical phonon resonances leading to strong spectral gaps in the emission. In this case, spintronic structures as novel THz sources may present real advantages in view of a broader THz emission window working at room temperature without spectral gaps~\cite{Seifert_2016}. Indeed, since the pioneering experiments of Beaurepaire \textit{et al.} in 1996~\cite{Beaurepaire1996}, followed by complementary investigations~\cite{Koopmans2000,Beaurepaire2004}, it is now well established that the excitation of a magnetic material by an ultrafast laser pulse leads to a picosecond demagnetization of ferromagnetic films within the 3\textit{d} band. Such demagnetization process is accompanied by ultrafast hot carrier spin-currents in the 4-\textit{sp} band, as proposed in MgO/Fe/Au multilayers~\cite{melnikov2011}. Recently, it has been demonstrated that spintronic elements composed of a ferromagnet (FM)/heavy metal (HM) bilayer can then emit relatively strong terahertz pulses owing to the conversion power of an excited spin current into a transient dipole charges~\cite{Huisman2015b,Jin2015,Huisman2016,Seifert_2016,Yang2016}. From quantum mechanic rules, THz emitters also represent an alternative in the perspective to reach a high degree of linear polarization (>95\%).

\vspace{0.1in}

In the context of spintronics, ultrafast spin-to-charge conversion (SCC) is responsible for the emission of linearly-polarized THz electromagnetic dipolar radiation. SCC is made possible owing to the so-called spin-orbit interactions (SOI) near or at interfaces between a FM and an active material characterized by a large atomic number $Z$ with strong SOI such 4\textit{d} (Pd) or 5\textit{d} (Pt, W, Ta, Au:Pt, Au:W, Au:Ta) HMs. Investigated extensively in the quasi-static excitation regime using ferromagnetic resonance (FMR) and spin-pumping (SP) techniques, SCC can occur \textit{via} the inverse spin Hall effect (ISHE)~\cite{saitoh2006,Hoffmann2013SpinHE,Sinova2015,Guo2008,Tanaka2008} owing to an asymmetric deflection of the spin trajectory induced by SOIs. Moreover, SCC may also arise from the inverse Edelstein-Rashba effect (IREE) that occurs at spin-orbit split Rashba interfaces states~\cite{EDELSTEIN1990233}. Such conversion was recently experimentally demonstrated on Rashba surface states~\cite{Sanchez2013,Zhou2018,Hoffmann2018} and on surface states of topological insulators (TI) in the THz domain~\cite{Hyunsoo2018}. Note that THz pulses from spintronic emitters only weakly depend on the circular polarization of the excitation laser beam, as observed from the optical selection rules in photogalvanic experiments~\cite{Huisman2015b}. In contrast, the phase of the emitted THz radiation and E-field may be controlled with the magnetization direction~\cite{Hibberd2019,Hoffmann2018,Hyunsoo2018} and permitting specific E-field waveform emission~\cite{wang2019}. This property is particularly appealing when considering an electrical control of the spin-based THz source output. Moreover, ultrafast SCC emitters possess the advantage of being phonon-less, resulting in the absence of any THz spectral dips, even at room temperature, and allowing extremely large unperturbed emission bandwidths (larger than $20$~THz~\cite{Seifert_2016}) compared to other technologies based on femtosecond oscillators.

\vspace{0.1in}

In an international context~\cite{Huisman2015b,Jin2015,Huisman2016,Seifert_2016,Yang2016,Hyunsoo2017,Seifert2017,Seifert_2018,Torosyan2018,Qiu18,Cramer2018,Herapath2019,hyunsoo2018b}, HM and Rashba materials have shown large THz emissivity comparable and even larger to that realised using mature nonlinear crystals such as ZnTe. Such emission properties are generally probed using standard THz-TDS set-up. The FM/HM junction, where Fe/Pt or Co/Pt represent the best candidates, exploits the ISHE with the so-called spin Hall angle (SHA), $\theta_{SHE}=j_c/j_s$. This represents the ratio between the transverse charge current $j_c$ generated over the spin current $j_s$ source, as the main figure of merit of the local SCC. When summed over the active thickness of layers, SCC scales with a characteristic length $(\theta_{SHE} \times l_{sf})$, the product of the SHA with the spin-diffusion length $l_{sf}$ (SDL), that lies around 0.2~nm for Co/Pt TMs. In the time-domain, this length matches the total volume of the transient dipole charge oscillations responsible for the dipolar emission. Improvements of the THz emission has been demonstrated using \textit{i}) engineered spintronic multilayers for optimal energy absorption of the laser pulse~\cite{feng2018}, or \textit{ii}) by using constructive interferences of ultrafast spin-currents in FMs with generated photoconductive carriers in a semiconductor switch~\cite{hyunsoo2018b}, or by using stacks of active multilayers~\cite{Yang2016}.

Beyond, the aforementioned spintronics structures are being extensively investigated for their spin-orbit torque (SOT) properties required for magnetic commutation functionalities. Progress in these materials is concentrated in particular on interface engineering for the optimization of the spin transmission~\cite{cornell2019c,cornell2019b,cornell2019a,Lee2019,zhucornell2019}. This is in order to significantly reduce the interfacial spin-loss as shown in recent experiments~\cite{jaffres2014,berger2018} and more recently \textit{via} first-principles calculations~\cite{belashchenko2016,nikolic2018,Zeng2019,kelly2019,kelly2020}. The spin-transmission efficiency is generally determined by combining spin-pumping (SP) and SP-ISHE experiments in the ferromagnetic resonance (FMR) regime or alternatively by spin-transfer torque (STT-FMR) techniques revealing the SOT generated \textit{via} reciprocal effects to spin-pumping. The process efficiency scales now with the electronic transmission, characterized as the spin-mixing conductance of the considered interface~\cite{Liu2011,berger2018}. The combination of SP-ISHE and THz techniques thus offers the ability to analyze the anatomy of spin-current injection in TM based multilayers from which the electronic transmission appears to be the key physical parameter~\textcolor{black}{\cite{jaffres2014,parkin2015,berger2018,ohno2019,kamp2020}} as discussed very recently in the analysis of THz emission from spintronic emitters~\cite{nenno2019,battiato2020}.
Moreover, very recent experiments of SCC by FMR spin-pumping/ISHE methods foresees possible breakthrough for THz emission supplied by Rashba-states exhibiting IEE characteristic lengths of 6~nm~\cite{lesne2016}, thus largely exceeding the one of Co/Pt.

\vspace{0.1in}

In this paper, using THz-TDS, we demonstrate and model a sizable spin-charge conversion and subsequent THz emission from transition metal (NiFe,Co)/Pt bilayers in the sub-picosecond timescale. The THz emission can be controlled by the direction of the magnetization in the plane of the FM layer revealing the ISHE origin. In particular, we extensively use the FMR spin-pumping/ISHE method to correlate the spin-injection temporal dynamics to the steady-regime of spin-injection. We model such mechanisms by advanced ab-initio methods and simplified Finite Difference Time-Domain simulations (FDTD). We show that, unlike same 3\textit{d}/5\textit{d} NiFe/Au:W and NiFe/Au:Ta systems showing poor electronic transmission at interface, the strong benefit of Co/Pt lies in an optimized electronic transmission for the majority spin-channel associated to a significant interfacial spin-asymmetry. By our numerical analysis method, we discuss the main issues giving rise to THz emission spectra and compare the main trends to more conventional spin-injection and spin-pumping experiments.

\textcolor{black}{In particular, one aim of the paper is to point out the strong correlation between the THz emission power of a given system to the efficiency of spin-charge conversion obtained in radio frequency (rf) spin-pumping experiments. How this correlation occurs? The THz emission power (in unit of the electric field $E_{THz}$) is proportional to an overall figure of merit $\eta\propto  g^{\uparrow\downarrow} \times \sigma_{SHE}  \times l_{sf}^{SOC}$ (in the case of ISHE process) displaying sequential electronic events and described by the product of the \textit{i}) the spin-mixing conductance (SMC) $g^{\uparrow\downarrow}$ connected to the interface electronic transmission, and like demonstrated by our numerical analyses by \textit{ii}) the spin-Hall conductivity responsible of the spin-charge conversion, and by \textit{iii}) the spin-diffusion length $l_{sf}$ in the SOC material representing the volume of spin and charge relaxation. Note that, in that picture, $g^{\uparrow\downarrow}$ represents the physical quantity probed also by FMR methods, except that the energy electron may differ by some 100~meV. Importantly, ab-initio calculations like presented in this contribution may answer that issue, also in order to optimize THz spintronic devices.}

The paper is organized into four different sections. \textcolor{black}{Section (\ref{Sec2}) is devoted to the description of THz-TDS spectroscopy method, of the different samples properties and resulting experimental data.} Section (\ref{Sec3}) is devoted to the characterization of transition-metal based systems by ferromagnetic resonance and spin-pumping-ISHE measurements in order to extract experimentally the electronic transmission at the relevant interfaces \textit{via} ISHE and SCC processes. Section (\ref{Sec4}) deals with ab-initio electronic structure and electronic transmission coefficient calculations of Co/Pt and NiFe/Pt bilayers, linking the high THz emission efficiency to hot electron excitation, carrier diffusion and spin-relaxation within spintronic bilayers and multilayers. \textcolor{black}{The final section (\ref{Sec5}) presents the results of the numerical simulations, analyses and interpretations before addressing the main conclusions.}

\section{\label{Sec2} Ultrafast spin-currents and spin-to-charge conversion probed by THz-TDS.}

\begin{figure*}[!htb]
\includegraphics[width=0.9\textwidth]{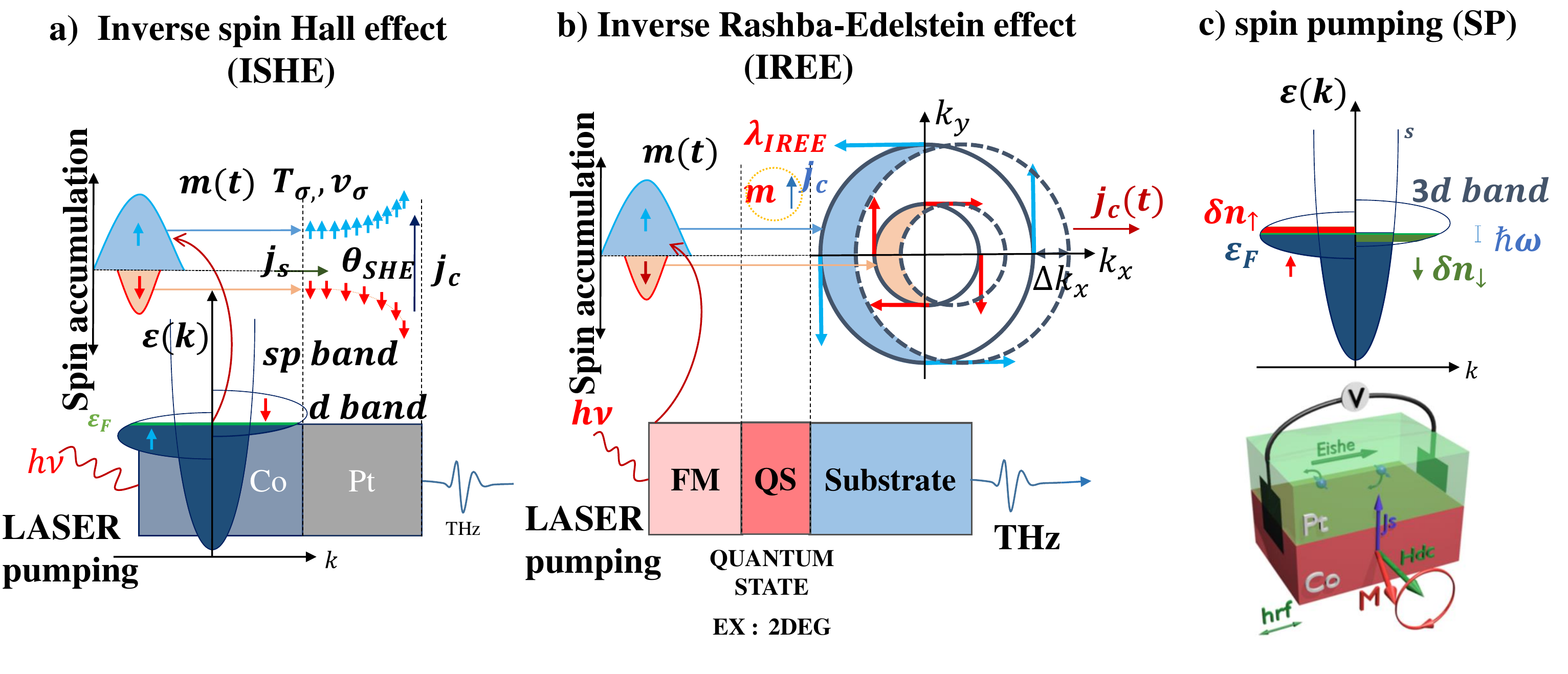}
\caption{Sketch representing the principles of operation of the three different experiments performed on spintronics interfaces that result in SCC through (a) laser induced inverse spin Hall effect (ISHE), (b) inverse Rashba-Edelstein effect (IREE) and (c) spin-pumping experiment. In the first two experiments, a short (100 fs) laser pulse excites hot spin-polarized electrons in the 4-\textit{sp} conduction band diffusing out after some energy loss into the adjacent layer where SCC will occur a) close to the interface by ISHE process or b) \textit{via} IREE process. c) In the spin-pumping experiment i.e. FMR-SP/ISHE, the out-of-equilibrium spin-density and spin-current are generated in the GHz regime in an energy range close to the Fermi level.}
\label{THz-TDS2}
\end{figure*}

THz emission from spintronic structures has been recently suggested to critically depend on the material properties like spin Hall angle and conductivity~\textcolor{black}{\cite{nenno2019}}. In that section, we demonstrate how, for materials and active interfaces with equivalent properties, the interface qualities and electronic transmission are critical parameters for THz emission~\cite{battiato2020}. To this end, we specifically discuss and emphasize on the experimental results of THz emission obtained on various spintronic Pt~\cite{jaffres2014,rojasapl}, Au:W~\cite{laczkowski2014,laczkowski2015} and Au:Ta~\cite{laczkowski2017}-based samples investigated by THz-TDS. \textcolor{black}{We compare such experiments, data and analyses to complementary RF-spin-pumping/ISHE results in the same series of samples (section II).} Indeed, the main SHE/ISHE properties of these materials and interfaces has been previously largely investigated in the steady-state regime of spin injection by radio-frequency (rf) spin-pumping. They will be largely discussed now in terms of ultrafast spin currents and subsequent conversion into transient charge current and THz emission. Co/Pt constitutes the actual reference in THz spintronic emitters. \textcolor{black}{In particular, from an electronic point-of-view, we have previously shown that Pt and Au:(W,Ta) materials differ in that their SHE/ISHE properties originate from the \textit{intrinsic} nature of their spin-Hall conductivity in the case of Pt~\cite{Tanaka2008,Guo2008,Sinova2015}, while originating from the \textit{extrinsic} side-jump effect for Au host-based alloys~\cite{laczkowski2017}.} It results in an expected enhanced electronic scattering and a larger resistivity for disordered alloys that could give rise to an increase of the effective spin Hall angle, as \textcolor{black}{also} evidenced recently in the case of Pt~\cite{cornell2019a,cornell2019b,cornell2019c,back2016,ywu2019}. It may also be \textcolor{black}{accompanied by} an increase of the side-jump effect in the case of Au based materials as discussed here. Moreover, we recall the general symmetry rules obeyed by the ISHE (\textit{resp.} SHE) phenomenon. ISHE can be described by a cross product between the three following vectorial quantities: the direction of the carrier spin $\bm{\sigma}$, the direction of the spin-polarized carrier flow corresponding to the spin-current $\bm{j}_s$ and the outgoing charge current $\bm{j}_c$ according to the following relationship $\bm{j}_c \propto \bm{\sigma} \times \bm{j}_s$ (\textit{resp.} $\bm{j}_s \propto \bm{\sigma} \times \bm{j}_c$ for SHE by Onsager reciprocal effects).

\subsection{\label{}Principle and THz-TDS set-up in the reflection mode.}

We briefly describe the experimental protocol and physical principles leading to the generation of the terahertz waves from spintronic emitters. Fig.~\ref{THz-TDS2} shows sketches of the three major experiments we focus \textcolor{black}{more particularly here} on: \textit{i}) the ultrafast spin-charge interconversion by ISHE \textcolor{black}{after excitation by a short laser pulse}, \textit{ii}) the ultrafast spin-charge interconversion by IEE on Rashba-states~\cite{Hoffmann2018,MoS2} after excitation by a short laser pulse, and \textit{iii}) the spin-pumping/ISHE \textit{via} FMR methods. Regarding ISHE, we typically consider the Co/Pt structure as pictured on Fig.~\ref{THz-TDS2}a). Initially, the short laser pulse generates hot spin-polarized carriers by absorption, in particular inside the ferromagnetic layer (Co), that diffuse to the Pt heavy metal (SOC material) after a short delay. ISHE may be considered as \textit{bulk}-type in the sense that SCC occurs in the Pt layer at the interface with Co over a typical length scales given by the spin-diffusion length of the material (even shorter of the order of some nm). Penetrating into the heavy metal, the ultrafast spin-current $\bm{j}_s=\bm{j}_\uparrow-\bm{j}_\downarrow$ gives rise to an ultrafast transverse charge current with a corresponding ratio $\bm{j}_c/\bm{j}_s$ given by the intrinsic spin Hall angle or SHA ($\theta_{SHE}$). This effect is \textit{bulk} also in the sense that the conversion is only possible from Bloch propagating electronic states. Although not the case for these bulk materials studied here, THz pulses can also be generated \textit{via} the Inverse Edelstein Effect (IREE, case \textit{ii} in Fig.~\ref{THz-TDS2}) at Rashba interfaces between ferromagnetic and two dimensional materials or surface quantum states (e.g. topological insulators). IREE also results in SCC for THz generation, where a spin-accumulation at the Rashba interface generates an in-plane electric field transformed into a charge current~\cite{Hoffmann2018,Qi18,MoS2}, phenomenon that we will not consider henceforth in the present paper.

\begin{figure}[!htb]
\includegraphics[width=0.7\columnwidth]{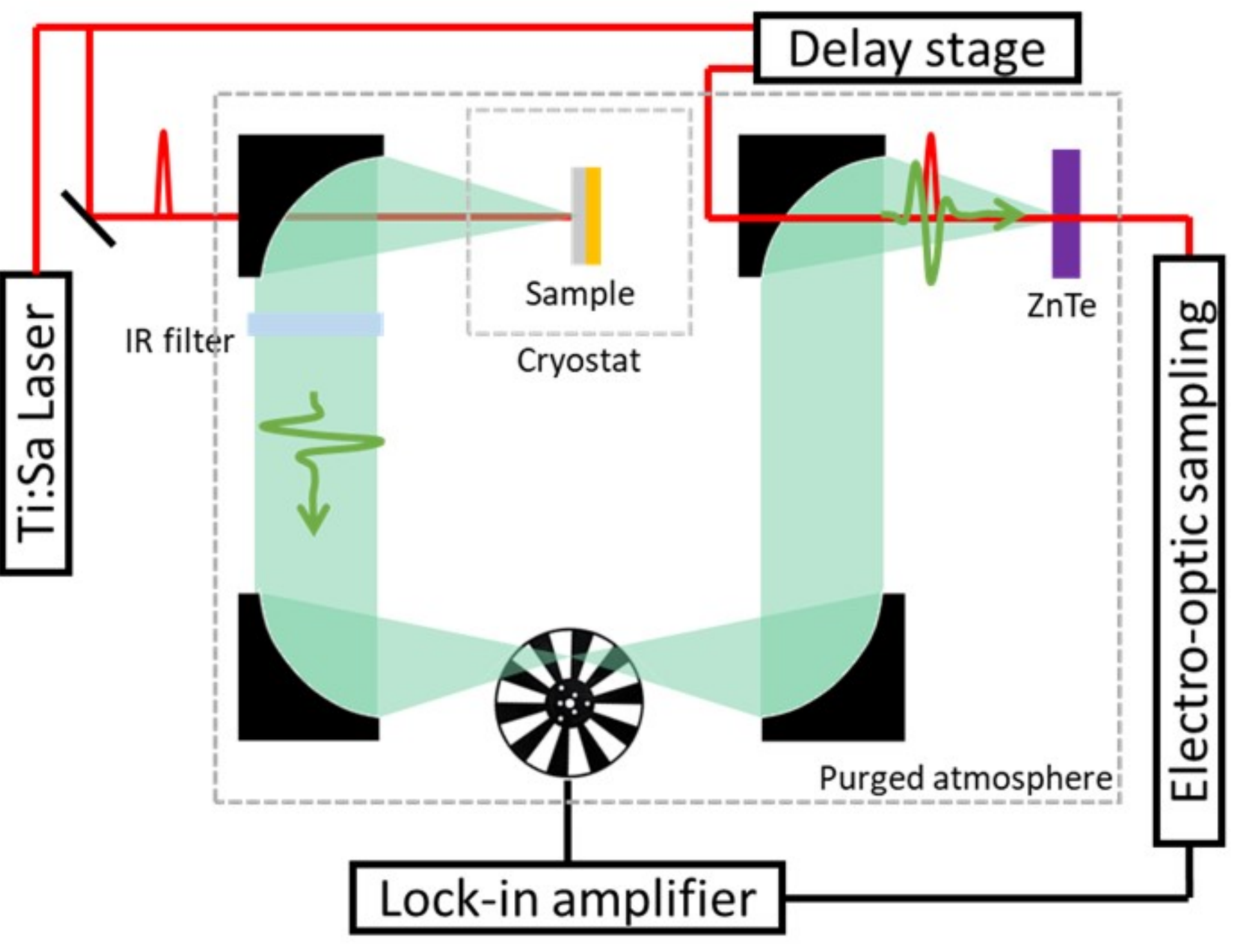}
\caption{\textcolor{black}{Sketch of the THz-TDS experimental setup used in reflection mode to characterize spin-emitters. A standard Ti:Sa oscillator is used to generate 100fs pulses with a repetition rate of 80 MHz centered at 800nm. This beam is used to induce THz radiation in the spintronic emitter and to pump the non-linear crystal (ZnTe) for electro-optic sampling. The THz generated is mechanically chopped at 6 kHz to enable a locking detection. The time profile of the THz electric-field is then sampled by moving the delay line. More details are given in the text.}}
\label{THzTDSSetup}
\end{figure}

The THz-TDS experimental setup is shown in Fig.~\ref{THzTDSSetup}. The emitters were placed on a mount with a small magnetic field (between $10$~mT and $0.2$~T) parallel to the spin interface. An ultrafast ($\simeq 100$ fs pulses) Ti:Sapphire oscillator centered at 810~nm was used to photo-excite the spin carriers directly from the front surface. Average powers of up to $\simeq 600$~mW were used \textcolor{black}{with a repetition rate of 80~MHz. The typical laser spot size on the sample was about $200\mu m\times 200 \mu m$.} The generated THz pulses were also collected from the front surface of the spin-emitter (\textit{i.e.} reflection geometry) using a set of parabolic mirrors \textcolor{black}{of respective 150~mm and 75~mm focal length to focus on samples.} Standard electro-optic sampling was used to detect the electric field of the THz pulses, using a 500~\text{$\mu$}m thick $\left<110\right>$ ZnTe crystal. A chopper is placed at the focal point between the second and third parabolic mirror to modulate the THz beam at 6 kHz for lock-in detection. A mechanical delay line is used to sample the THz ultrafast pulse as a function of time. The THz-TDS setup is placed in a dry-air purged chamber (typically $<2$\% humidity) to reduce water absorption of THz radiation.

\subsection{\label{}Samples preparation and characterization - Main properties.}

The first set of samples are made of glass//Co(2)/Pt(4) and glass//NiFe(2)/Pt(4) 3\textit{d}/5\textit{d} transition-metal based bilayers where the numbers in bracket indicates the thickness in nm. The structures are grown by magnetron sputtering in a single deposition chamber on borosilicate glass to avoid large backward THz absorption. \textcolor{black}{The layer thickness was fixed after having calibrating the deposition speed on thicker films \textit{via} X-ray reflectivity measurements.} These samples constitute reference THz emitters characterized by emission power equivalent or larger than those obtained on ZnTe or GaP nonlinear crystals. A second series of 3\textit{d}/5\textit{d} transition metal samples are made of NiFe(2)/Au:W$_{0.13}$(4) and NiFe(2)/Au:Ta$_{0.05}$(4) structures deposited by co-sputtering (co-deposition) of the two pure materials. In (Co or NiFe)/Pt devices, Pt is known to provide an intrinsic spin-Hall effect (SHE) mechanism which is scaled by an effective spin-Hall conductivity \textcolor{black}{ ($\sigma_{SHE}^{Pt}$) of $\sigma_{SHE}^{Pt}=3.2\times 10^3\Omega^{-1}$.cm$^{-1}$~\cite{Guo2008,Tanaka2008} and a subsequent effective spin Hall angle $\theta_{eff}^{Pt}=\rho_{xx}\sigma_{SHE}$ of about $0.05$ in the in-plane current geometry (spin-current perpendicular to plane) owing to its resistivity, $\rho_{Pt}=17~\mu \Omega$.cm of Pt at room temperature. Those physical parameters were determined \textit{via} combined FMR-spin-pumping ISHE experiments on varying the Pt layer thickness~\cite{jaffres2014}.
On the other hand, Au:W~\textcolor{black}{\cite{laczkowski2014,laczkowski2015}} as well as Au:Ta~\textcolor{black}{\cite{laczkowski2017}} impurity alloys are known to promote an extrinsic \textit{side-jump} spin-Hall effect depending on the impurity content. The corresponding value may reach $\theta_{eff}^{Au:W}=0.15$ for Au:W and $\theta_{eff}^{Au:Ta}=0.4$ in the best alloys. The determination of those parameters were performed by using combined FMR-spin-pumping/ISHE measurements as well as possibly non-local spin-injection/detection in lateral spin-valve (LSV) devices when possible. The readers are invited to refer to the following references~\cite{laczkowski2014,laczkowski2017} for details.}
In the present case, Au:W$_{0.135}$ and Au:Ta$_{0.05}$, we get respectively $\theta_{eff}^{Au:W}=0.15$ and $\theta_{eff}^{Au:Ta}=0.25$ corresponding to a spin-Hall angle larger than that of Pt. However, as expected, a much larger resistivity $\rho_{Au:W_{0.13}}=90~\mu \Omega$.cm and $\rho_{Au:Ta_{0.05}}=50~\mu \Omega$.cm compared to that of Pt $\rho_{Pt}=17~\mu \Omega$.cm is measured at room temperature for that alloys. \textcolor{black}{Material resistivity were measured \textit{via} four contacts Van der Pauw methods}.

\subsection{\label{}THz-TDS of ultrafast spin-currents.}

\begin{figure}[htb]
\includegraphics[width=0.9\columnwidth]{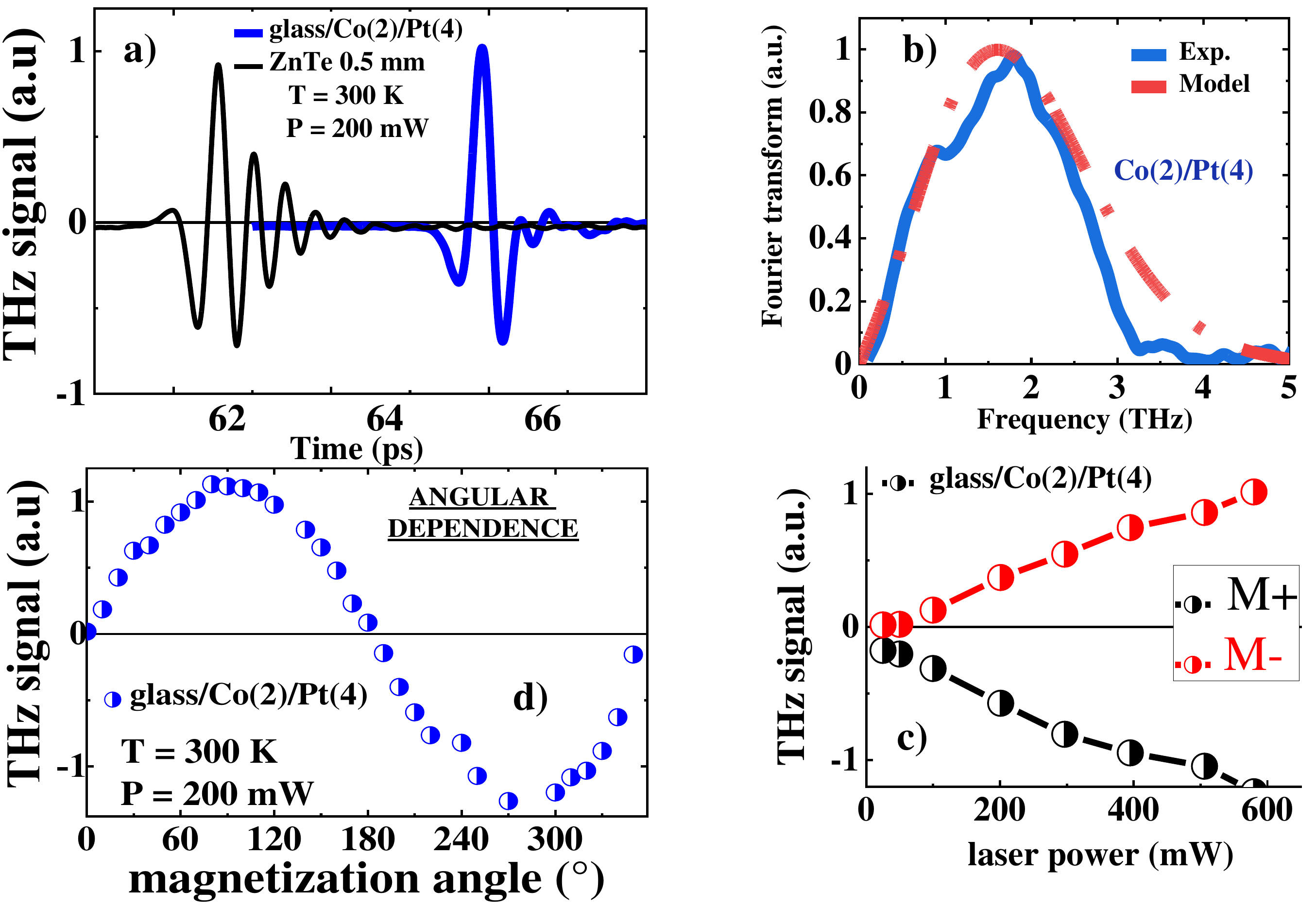}
\caption{a) Temporal THz emission acquired using THz-TDS at room temperature on Co(2)/Pt(4) system under a 100 fs laser pulse with average power of $P=200$~mW. Comparison with THz emission from ZnTe crystal \textit{via} optical rectification under the same experimental conditions. b) \textcolor{black}{Experimental Fourier transform (FT) of the Co(2)/Pt(4) temporal signal giving the spectrum in the $[0-3]$ THz window (blue).} Comparison with the modelling using parameters described in the text (black dashed line).  c) Amplitude of the maximum THz wave vs. laser power excitation in the $[0-600]$~mW range, \textcolor{black}{repetition rate of 80~MHz} and under two different opposite magnetic polarities. d) Characteristic sine angular dependence of the THz-wave amplitude by varying the in-plane Co magnetization angle.}
\label{THz-Co}
\end{figure}

We describe the THz emission from the 3\textit{d}/5\textit{d} transition metal HM/FM elements described above. Fig.~\ref{THz-Co}(a) displays the typical emitted electric field as a function of time for the Co/Pt sample at room temperature under an \textcolor{black}{medium} excitation power of 200~mW. \textcolor{black}{Concerning the whole experiments, we have not observed particular strong heating of the different metallic samples mainly due to the fact the absorption of the laser power is limited to less than 20\% for transition metal thin films. Moreover, experiments and THz signal shows a strong reproducibility in time on the exact same sample series.} Generally, THz signal highlights the relaxation in Pt of a hot electron spin current, formed by spin $\uparrow$ and spin $\downarrow$ populations, that is converted to a transient charge current by ISHE. This displays the typical linear slope crossing the origin and representative of the derivative of the transient ultrafast charge current from SCC with $E_{THZ}\propto \frac{\partial J_c(t)}{\partial t}$ \textcolor{black}{(like more detailed hereafter}). The typical THz emission is comparable to that from a 0.5~mm thick ZnTe crystal owing to optical rectification in the exact same experimental excitation conditions \textcolor{black}{given in the previous subsection}. (Similar THz pulses are obtained on Co/Pt samples grown on high-resistive (HR) Si substrates \textcolor{black}{of about the same magnitude}). \textcolor{black}{Fig.~\ref{THz-Co}b) shows the typical experimental spectrum of the spintronic emitter obtained by Fourier Transform of the temporal emission in the $[0-3]$ THz window, limited by our detection setup. From literature~\cite{Seifert_2016}, spintronic THz emitters are known to cover a wider emission range up to 30~THz.}

\textcolor{black}{We anticipate now the discussion about the modelling of such spectra which will be discussed more in details in section IV and give the main issues. Fig.~\ref{THz-Co}(b) black dashed line displays the results of our modelling when one considers the spin-charge conversion into Pt in the time domain after an excitation pulse in Co. This has been derived from the spin-dependent Maxwell equation in a bilayer (FM/HM) taking into account high electronic transmission at interfaces (section IV) and high conductivity in the SOC material (Pt). We manage to reproduce conveniently the FT specta by considering momentum $\tau$ and spin-relaxation time $\tau_{sf}$ for Co and Pt extracted from the literature for energy not too far from the Fermi energy (refer to Table 1 for physical parameters used). Indeed, the high electronic transmissivity at Co/Pt interface for a large energy window above the Fermi level makes less critical the energy-dependence of the carrier relaxation and electron flow does not discard too much from a diffusive picture, before their fast relaxation in Pt. When the transmission is expected to vary much with energy and for a rather slow relaxation in the NM (\textit{e.~g}), a superdiffusive picture (not considered here) is necessary. Departing from this Co/Pt system, scaling up the relaxation rate of hot electrons, that is decreasing in proportion the momentum and spin relaxation time, is accompanied by an enhancement of the resistivity (case of NiFe/Au:W and NiFe/Au:Ta) and a loss of the THz signal due to a larger current dissipation. We give some modelling and simulation results in section IV.}

Fig.~\ref{THz-Co}(c) demonstrates the variation of the electric field as a function of power and shows two important features. First, it demonstrates that the amplitude of the THz waveform increases almost linearly with the laser excitation power within the range $[0-600]$~mW. It will be shown to be very analog to the output transverse ISHE voltage \textit{vs.} rf-power in more conventional FMR-spin pumping-ISHE experiments. Second, the sign or polarity of the E-wave THz field can be reversed upon magnetization reversal along two opposite directions. This is a particular feature of the THz emission from spin current injection and this originates from the direct quantum symmetry rules of the SHE/ISHE as discussed above. Indeed, Fig.~\ref{THz-Co}d) shows the typical angular dependence of the THz electric field when the applied magnetization is rotated in the film-plane over the full $[0-360]$ degrees range. We can observe an almost perfect sine angular variation where the angle $\theta_M=0$ corresponds to the direction of the ZnTe crystal proper axis (direction of the analyzer). This exhibits a maximum of the signal when the magnetization is perpendicular to the detector \textcolor{black}{principal} axis or equivalently, that the electric field is perpendicular to the magnetization direction according to the SHE/ISHE conversion rules.

\begin{figure}[!htb]
\includegraphics[width=0.9\textwidth]{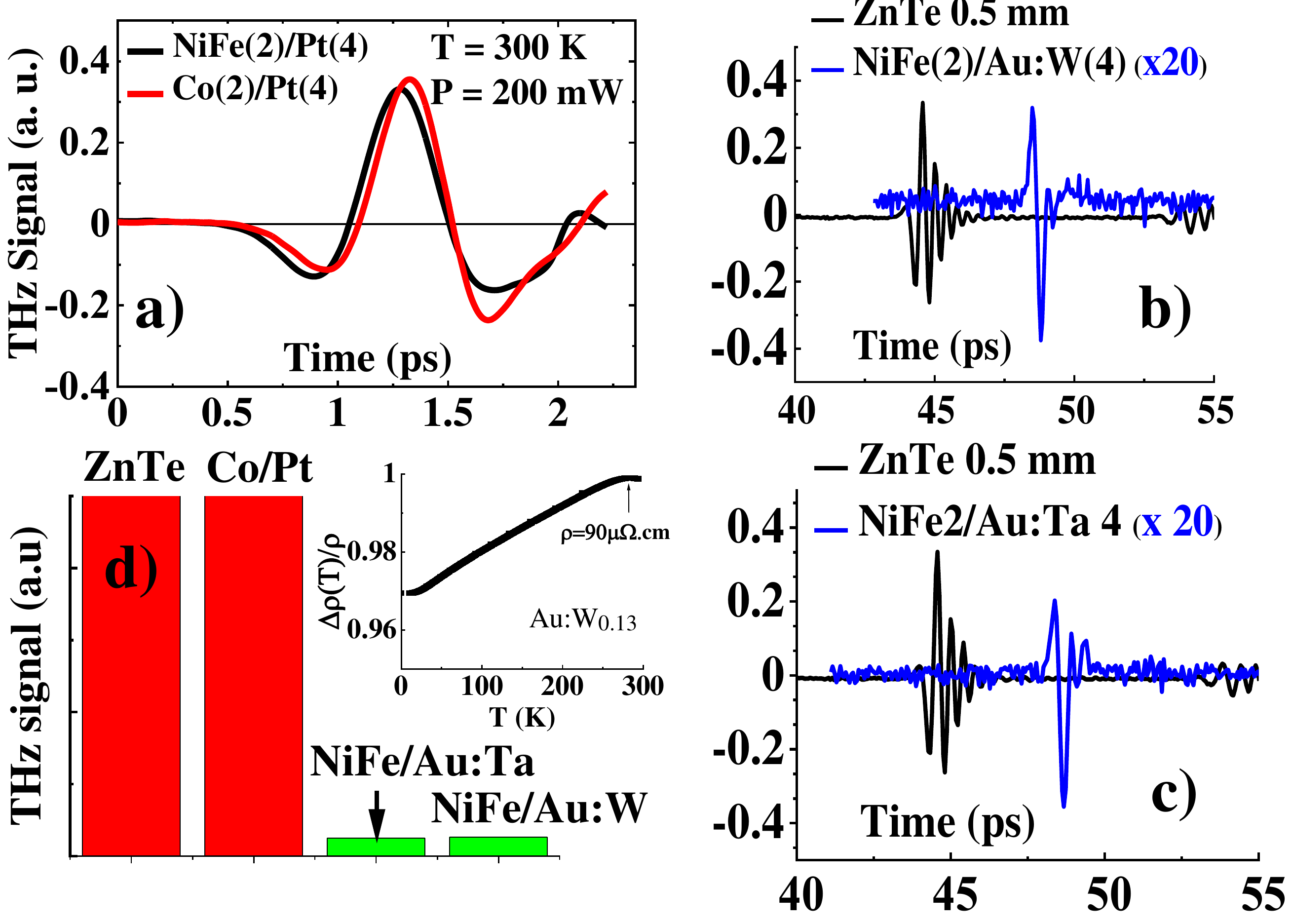}
\caption{a) Comparison of the THz emission at room temperature ($P=200$~mW) acquired on Co(2)/Pt(4) and NiFe(2)/Pt(4) samples showing similar features and equivalent amplitudes. b) c) Emission of NiFe/Au:(W,Ta) systems compared with ZnTe(110) crystal. Measurements taken at room temperature under a laser fluence of $P=200$ mW d) Peak-to-peak amplitude comparison between the well-known THz emitter ZnTe(110) and spintronic-based emitters including the reference Co/Pt and NiFe/Au(W,Ta) systems. Inset of d) Temperature variation of the resistivity $\frac{\Delta \rho_{Au:W_{0.13}}}{\rho}$ of Au:W$_{0.13}$ from 300 to 10~K showing a small decrease of 3\%. The sample Au:W$_{0.13}$ corresponds to a typical resitivity of $\rho\simeq 90 \mu\Omega.$cm at RT for an approximate W content of about 13\%.}
\label{FigTHzCo2}
\end{figure}

Fig.~\ref{FigTHzCo2}(a) displays the comparison of the THz-TDS spectra acquired between Co/Pt and NiFe/Pt samples at room temperature in the exact same excitation configuration. We observe an almost identical response in both cases, and this is stark contrast to demagnetisation experiments of the 3\textit{d} magnetic band~\cite{gorchon2017}. Our experiments shows the equivalence of the two systems involving Co or NiFe as a ferromagnetic source when one considers, not the dynamics of the local 3\textit{d} magnetization \textcolor{black}{like in Ref.~\cite{gorchon2017}}, but those of hot electrons in the 4\textit{sp} bands \textcolor{black}{location of mobile carriers}. \textcolor{black}{Moreover, one has performed equivalent experiments on glass//Co(2)/Al(2)/Al$_{ox}(3)$ and glass//NiFe(2)/Al(2)/Al$_{ox}(3)$ reference samples free of any heavy-metal layer. Results (not shown) display a signal for Co(2)/Al(2) of the order of 1/20 compared to Co(2)/Pt(4) and no observable signal for NiFe(2)/Al(2) giving thus the efficiency of self-emission (self-ISHE) for Co like reported recently~\cite{wu2019}.} One can conclude that the optimized performances obtained with 3\textit{d}/Pt systems clearly identify Co/Pt and NiFe/Pt as the \textcolor{black}{one of the most relevant interfaces} owing to their optimized spin-dependent electronic transmission. This will be shown theoretically by advanced ab-initio methods in Section \ref{Sec4}. We emphasize here the impact of two linked and key physical parameters for NiFe/Au:W and NiFe/Au:Ta systems on the THz emission, namely \textit{i}) a larger spin Hall angle $\theta_{SHE}$ compared to Co/Pt reference (respectively $0.15$ for Au:W and $0.25$ for Au:Ta) but with also a \textit{ii}) larger alloying resistivity $\rho$ due to scattering enhancement. Fig.~\ref{FigTHzCo2}(b) (Au:W) and Fig.~\ref{FigTHzCo2}(c) (Au:Ta) display the typical THz emission for each case, showing the typical waveform of hot spin current relaxation at the ps scale but with a strong reduction in field by a factor 20 compared to Co/Pt or NiFe/Pt. This observation is surprising as an increase by a factor between 2 and 5 would be expected owing to the larger SHA in the Au-based alloy systems. The results are summarised on Fig.~\ref{FigTHzCo2}(d). As demonstrated below, the reason for such a decrease are potentially twofold: \textit{i}) the enhancement of the alloy resistivity strongly decreases the THz emission due to spin current relaxation effects and \textit{ii}) a reduction of the spin-transmission at NiFe/Au interfaces  compared to Co/Pt or NiFe/Pt, \textcolor{black}{as revealed by FMR-spin-pumping experiments}.

\section{\label{Sec3}Spin-pumping experiments and interface transparencies.}

The second type of experiment is FMR spin-pumping (FMR-SP) coupled to ISHE, which can be performed on the same type of structures~\cite{saitoh2006}. FMR-SP (Fig.~\ref{THz-TDS2}c) consists in generating at the FMR resonance,in the typical range $6-10$~GHz an out-of-equilibrium spin-current. SCC occurs as described above except that the out-of-equilibrium spin-polarized carriers remains within an energy range $\epsilon$ close to the Fermi energy $\epsilon_F$ $(\epsilon\simeq \epsilon_F\pm \hbar\omega)$ leading to a small DC current or voltage. This out-of-equilibrium spin-current is defined as:

\textcolor{black}{
\begin{eqnarray}
j^{eff}_s &= &\frac{2e^2}{\hbar}g_{eff}^{\uparrow\downarrow} \frac{\hbar \omega}{e}sin^2\left(\theta_m\right)\propto P_{rf}\nonumber\\
j^{eff}_s &= &\left(\frac{2e}{\hbar}\right)\frac{g_{eff}^{\uparrow\downarrow} \gamma^2\hbar h_{rf}^2}{8\pi \alpha^2}\frac{4\pi\gamma M_{eff}+\sqrt{\left(4\pi\gamma M_{eff}\right)^2+4\omega^2}}{\left(4\pi\gamma M_{eff}\right)^2+4\omega^2}
\end{eqnarray}}
where $e$ is the electron charge, $\hbar$ is the reduced Planck constant, $\omega=2\pi f$ is the pulsation of the rf-field and $f$ the rf-frequency, $\gamma=g\mu_B/\hbar$ is the gyromagnetic ratio with $g$ the Landé factor and $\mu_B$ the Bohr magneton, $M_{eff}$ is the effective saturation magnetization, \textcolor{black}{and $h_{rf}$ the small oscillating rf transverse field responsible for precession}. \textcolor{black}{$\theta_m$ stands then for the average precessing angle of the magnetization around the local effective static field $H_{DC}$ with the result that $sin^2\left(\theta_m\right)$ is proportional to the rf-power absorbed by the ferromagnetic material. It results that the charge current pumped after spin-charge interconversion in the heavy metal is directly proportional to $P_{rf}\propto sin^2\left(\theta_m\right)$ and then to $h_{rf}^2$; and that a normalized value for it writes in $A/G^2$}.

The term \textit{eff.} corresponds to a  \textit{effective} renormalization compared to the maximum spin-current that can be injected and takes into account \textit{i}) the finite electron transparency at the interface, and \textit{ii}) a possible spin-backflow from the HM towards the FM. Note that in the previous expression, one has $\alpha=\alpha_0+\Delta \alpha$ where $\alpha_0$ is the bare damping free of spin-current dissipation and $\Delta \alpha$ the damping enhancement due to spin-pumping. The expression of the spin-current injection by the $\alpha^2$ factor in the denominator accounts for the reduction of the characteristic precession angle when $\alpha$ increases. Generally, the determination of the $\alpha$ parameter is performed by extracting the linear dependence and slope of the FMR linewidth \textit{vs.} the rf-frequency as performed previously in the case of the Co/Pt systems.  \textcolor{black}{$g_{eff}^{\uparrow\downarrow}=\sum_{k_{\parallel}} \left(1-r_\uparrow r_\downarrow^*\right)$ is the effective spin-mixing conductance per unit area} of the corresponding interface, as a sum of both real and imaginary parts of the aforementioned quantity over the incoming $k_\parallel$ channels. This non-zero quantity is thus responsible for the spin-current dissipation and FMR linewidth ($\alpha$) enhancement with the result that the damping parameter $\alpha$ may vary over a very short length scale, smaller than 2~nm. We give here the main rules describing the spin-current dissipation at interfaces from spin pumping data. The increase of the parameter $\alpha$ is generally related to $g_{eff}^{\uparrow\downarrow}$ by the following relationship:

\begin{equation}
   \Delta \alpha=\frac{g\mu_B}{4\pi M_{eff}t_{FM}}g_{eff}^{\uparrow\downarrow}
\end{equation}
where $t_{FM}$ is the thickness of the ferromagnetic layer. In this expression, $g_{eff}^{\uparrow\downarrow}$ is the effective spin-mixing conductance taking into account the spin-backflow process and spin-memory loss or SML~\cite{jaffres2014}. The decrease of the spin-injection efficiency may reach 60\% that way, and this loss of spin-injection efficiency can be treated by the so-called $R_{SML}$-coefficient~\cite{jaffres2014,nikolic2018,kelly2020}:

\begin{equation}
   R_{SML}=\frac{j_s^{HM}}{j_s^{eff}}=\frac{r_{sI}}{r_{sI}\cosh(\delta)+r_{s,\infty}^{HM}\coth\left(\frac{t_{HM}}{l_{sf}^{HM}}\right)\sinh(\delta)}
\end{equation}
where $j_s^{eff}$ and $j_s^{HM}$ are the spin current escaping the FM and the one injected in the HM which may now differ. $r_{sI}$ is the interfacial spin-resistance, ratio of the boundary resistance (interface resistance) $r_b$ is the resistivity of the interface and and $\delta$ the SML parameter. $r_{s,\infty}^{HM}=\rho_{HM} \times l_{sf}^{HM}$ is the spin-resistance of the heavy metal of thickness $t_{HM}$ and spin-diffusion length $l_{sf}^{HM}$. $R_{SML}$ appears as a \textit{constant} in the sense that it does not depend on the spin-flip rate in FM. We have:

\begin{widetext}
\centering
\begin{equation}
 g_{eff}^{\uparrow\downarrow}=g^{\uparrow\downarrow}\frac{r_{sI}\cosh(\delta)+r_{s}^{HM^\infty}\coth(\frac{t_{HM}}{l_{sf}^{HM}})\sinh(\delta)}{r_{sI}\left[1+\frac{1}{2}\sqrt{\frac{3}{\epsilon}}\coth(\frac{t_{HM}}{l_{sf}^{HM}})\right]\cosh(\delta)+\left[r_{s}^{HM^\infty}\coth(\frac{t_{HM}}{l_{sf}^{HM}})+\frac{1}{2}\frac{r_{sI}^2}{r_{s}^{HM^\infty}}\sqrt{\frac{3}{e}}\right]\sinh(\delta)}
 \label{SMC0}
 \end{equation}
\end{widetext}
where $\epsilon=\tau_{p}/\tau_{sf}^{HM}$ is the spin-orbit parameter (probability of spin-flip after diffusion). Typically $\epsilon\simeq 0.1$ for Pt with an even larger value for Au:W and Au:Ta transition-metal alloys. \textcolor{black}{SML and effective SMC, like discussed above, are very important issues in the field of spin-orbit torques and THz emission spectroscopy may represent a powerful method in the future to probe the exact anatomy of spin-current at spintronic interfaces like displayed in a recent literature.}

\subsection{Samples preparation and FMR spin-pumping set-up.}

We now turn to experimental data acquired on Pt~\cite{jaffres2014,jaffres2018} and Au-based alloys materials~\cite{laczkowski2014,laczkowski2015,laczkowski2017}. For these experiments, we prepared a series of thicker ferromagnetic and spin-orbit based material samples, respectively Co(15)/Pt(30), NiFe(15)/Pt(30) as well as NiFe(15)/Au:W$_{0.135}$(30) and NiFe(15)/Au:Ta$_{0.05}$(30) grown by magnetron sputtering in a single deposition chamber on SiO$_2$-terminated Si wafers and in the exact same conditions as the ones prepared on glass for the THz emission experiments. A thicker ferromagnetic layer (15~nm) is suitable to enhance to the FMR resonance, whereas the thicker SOC material is beneficial to avoid the electrical rectification contribution to the signal. \textcolor{black}{Indeed, in spin-pumping experiments, the generation of a charge current into Pt needs to avoid a too much lateral expansion in the ferromagnet where anisotropic magnetoresistance (AMR) and Planar Hall effect (PHE) may give rise to a spurious electrical signal of a well defined angular dependence~\cite{iguchi2016}}. Samples were then cut in an elongated rectangular shape of typical dimension $L\times W = 2.5\times 0.5$~mm$^2$. Combined FMR and inverse spin Hall effect (ISHE) measurements were performed at room temperature in a split-cylinder microwave resonant cavity. The radio-frequency (rf) magnetic field $h_{rf}$ was set along the long axis and the external applied DC magnetic field, $H_{dc}$ along the width of the rectangle. The frequency of $h_{rf}$ was fixed around $f=9.6-9.8$ GHz whereas $H_{dc}$ was swept through the FMR resonance condition. The strength of $h_{rf}$ was determined by measuring the quality factor ($Q$) of the resonant cavity with the sample placed inside for each measurement. At the resonance condition, the derivative of FMR energy loss was acquired as the voltage was recorded across the long extremity of the sample, that is along the perpendicular direction to the $H_{dc}$ field and magnetization. We have also performed a frequency dependence within the $3-24$ GHz range of the FMR spectrum in order to extract the effective saturation magnetization $M_{eff}$ as well as the damping constant $\alpha$. A refined analysis of the damping constant generally requires a reference sample free of spin-current dissipation at interfaces, \textit{i.e.} without spin-memory loss (SML). Ideally one would use a single Co layer capped by an oxidized Al capping layer.

\begin{figure*}[!htb]
\includegraphics[width=0.75\textwidth]{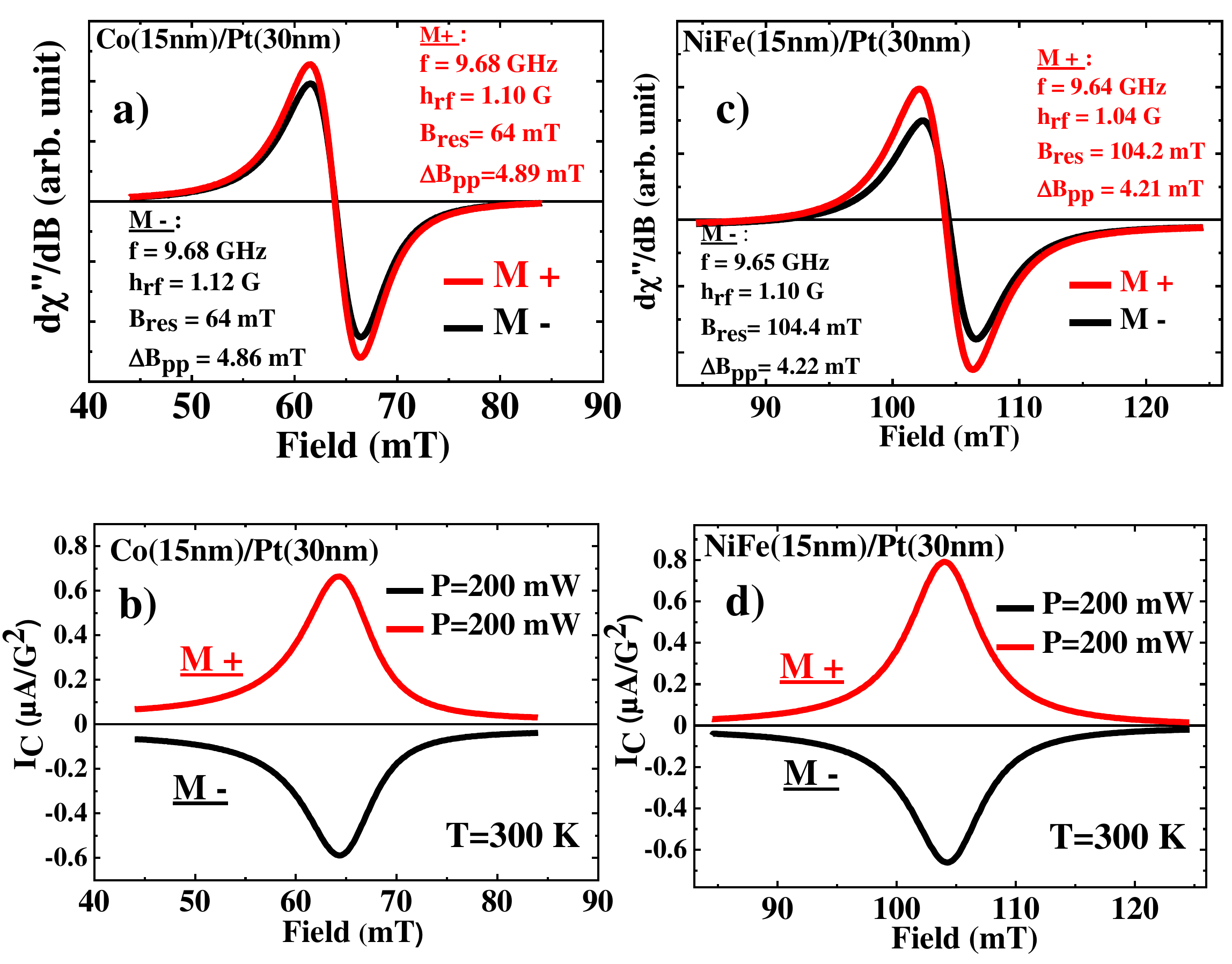}
\caption{Ferromagnetic resonance (FMR) acquired by the derivative of the absorption spectra for the respective positive ($M+$) and negative ($M-$) magnetization directions for (a) Co(15)/Pt(30) and (c) NiFe(15)/Pt(30) samples. The frequency is fixed at $f=9.68$ GHz and the rf-field is about \textcolor{black}{$h_{rf}=1.1$~G}. Corresponding ISHE current spectra acquired on Co(15)/Pt(30) and NiFe(15)/Pt(30), respectively (b) and (d), in the same experimental conditions and displaying the same quantitative and qualitative output signal.}
\label{FMR}
\end{figure*}

\subsection{Experimental results.}

We first focus on the results obtained on Co/Pt and NiFe/Pt characterized by a high electronic transparency and large THz emission. Fig.~\ref{FMR}(a,c) display the typical FMR spectra in the neighborhood of the resonance at frequency $f=9.8$ GHz by sweeping the dc in-plane magnetic field. These typical spectra, corresponding to the field derivative of the FMR absorption $\frac{d\chi^{\prime\prime}}{dH_{dc}}$, are almost unchanged when the magnetization is reversed (dc field) $+M\rightarrow -M$ in the cavity, are almost equivalent for the two samples made of Co or NiFe (Py). Fig.~\ref{FMR}(b,d) displays, in each case, the corresponding field dependence of the resulting transverse ISHE current $I_C$. This ISHE current level has been extracted by dividing the acquired transverse voltage by the device resistance ranging between 30 and 90 $\Omega$. In both cases of Co and NiFe-based samples, the level of current lies in the range $0.6-0.7~\mu \text{A.G}^{-2}$ which is a signature of \textit{i}) a very high carrier transmission efficiency and an efficient spin-charge interconversion for Pt with spin-Hall angle close to $\theta_{eff}^{Pt}\simeq 0.05$. The spin-Hall angle for Pt can be largely enhanced by CIP, alloying with a transition-metal species or synthesized in multilayered stacks~\cite{parkin2015,berger2018}. Moreover, when the dc external field and consequently the magnetization are reversed by $180^\text{o}$, the sign of the ISHE current $I_C$ changes polarity, as is expected by the SCC rule from the ISHE $\bm{j}_c \propto \bm{j}_s \times \hat{z}$ where $\hat{z}$ is the carrier flow direction. From Eq.~\eqref{SMC0}, this results in the spin-mixing conductance at saturation for Co/Pt is estimated as $g_{0}^{\uparrow\downarrow}\simeq 80 \times 10^{18}$ m$^{-2}$ giving rise to an effective spin-mixing conductance $g_{eff}^{\uparrow\downarrow}=\frac{g_{0}^{\uparrow\downarrow}}{1+g_{0}^{\uparrow\downarrow}r_{sI}}\simeq 22 \times 10^{18}$ m$^{-2}$ mainly limited by the SML resistance $r_{sI}$. A similar value of $g_{eff}^{\uparrow\downarrow}$ for NiFe/Pt is found, in agreement with the work of Berger~\cite{berger2018} \textit{et al}.

\begin{figure}[htb]
\includegraphics[width=0.7\textwidth]{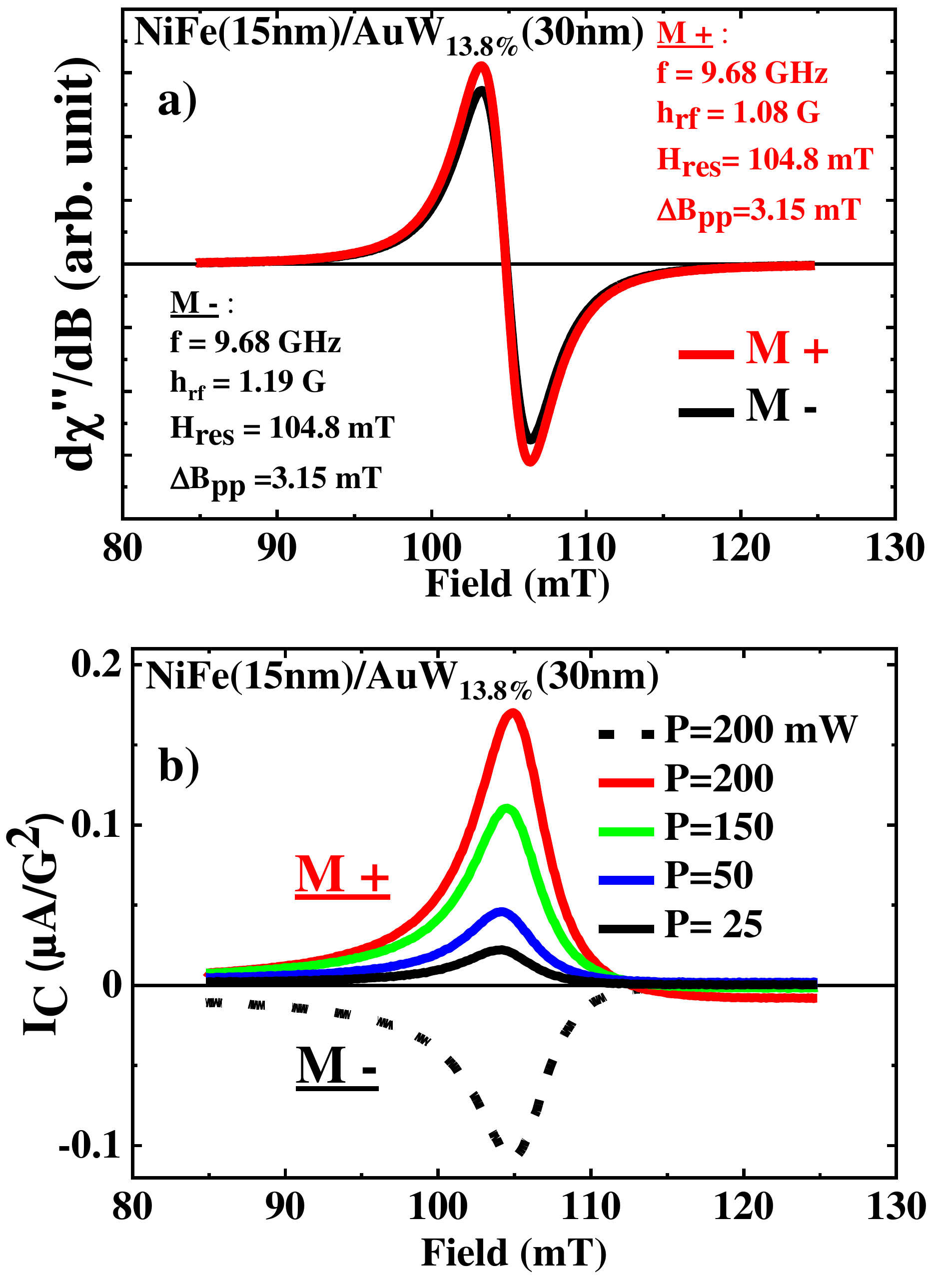}
\caption{a) Ferromagnetic resonance (FMR) acquired at room temperature by the derivative of the absorption spectra for positive ($M+$) and negative ($M-$) magnetization directions for NiFe(15)/Au:W$_{0.135}$(30). The frequency is fixed at $f=9.68$ GHz and the rf-field is about $B_{rf}=1.1$ Gauss. \textcolor{black}{The resonance field was found to 104.8 mT in each case. This emphasizes on the same absorption spectra acquired when switching the direction of magnetization.} b) Inverse Spin Hall effect (ISHE) current spectra acquired correspondingly \textcolor{black}{in the exact same experimental conditions and for different rf-power (from 20~mW to 200~mW). The sign inversion of the DC-current signal is representative of the ISHE effect. Note that the ISHE signal increases almost linearly with the rf-power injected.}}
\label{FMRSP}
\end{figure}

\begin{figure}[htb]
\includegraphics[width=0.7\textwidth]{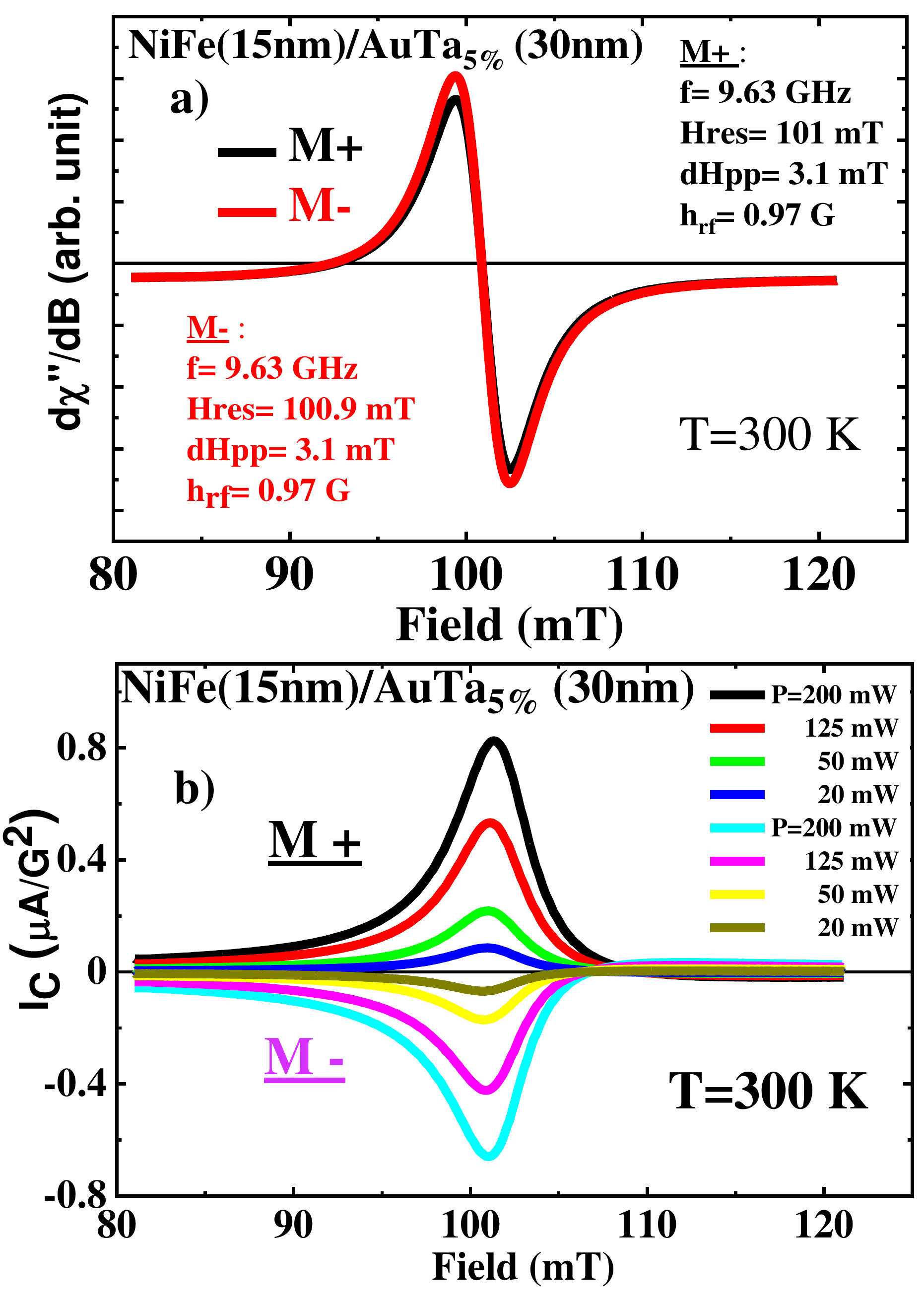}
\caption{a) Ferromagnetic resonance (FMR) acquired at room temperature by the derivative of the absorption spectra for positive ($M+$) and negative ($M-$) magnetization directions for NiFe(15)/Au:Ta$_{0.05}$(30). The frequency is fixed at $f=9.68$ GHz and the rf-field is about $B_{rf}=1$ Gauss. \textcolor{black}{The resonance field was found to 101 mT in each case. This emphasizes on the same absorption spectra acquired when switching the direction of magnetization. b) Inverse spin Hall effect (ISHE) current spectra acquired correspondingly in the exact same experimental conditions and for different rf-power (from 20~mW to 200~mW). The sign inversion of the DC-current signal is representative of the ISHE effect. Note that the ISHE signal increases almost linearly with the rf-power injected.}}
\label{FMRSPb}
\end{figure}

We now turn to the NiFe/Au:W and NiFe/Au:Ta sample series, \textcolor{black}{displayed respectively on Fig.~\ref{FMRSP} Fig.~\ref{FMRSPb}}, and characterized by a high spin-conversion efficiency (larger SHA) but modest THz emissivity. In these samples, the resistivity  for Au:W and Au:Ta, \textcolor{black}{ $\rho \simeq \frac{\hbar}{e^2}\left(\frac{2\pi}{k_F}\right)^2 N_D\sigma_{sc}$ measured by four-points Van Der Pauw method ($N_D$ is the impurity content, $k_F$ the Fermi wavevector and $\sigma_{sc}$ is the scattering cross section, $\lambda_{mfp}=N_D^{-1}\sigma_{sc}^{-1}$ is the mean free part insensitive to the temperature) is expected to be almost insensitive to phonon scattering and then to temperature due to the dominant role of impurities like measured previously on Au:W$_{0.07}$ where $\frac{\Delta \rho}{\rho}\simeq 3\%$ from 300 K to 10 K (see. Fig.~~\ref{FigTHzCo2}(d) in inset for the case of Au:W$_{0.13}$~\cite{laczkowski2014}. From, the previous expression it is possible to extract a rough value of the scattering cross section of W impurities in Au in the case of Au:W. If one considers average Fermi wavevector in Au of the order of $k_F=\frac{\pi}{2a}$ where $a$ is the lattice parameters, one obtains a pretty short distance $\frac{\sigma_{sc}}{a}\simeq1.5$\AA~thus corresponding then a strong impurity screening by the noble Au metal as a host.} We also expect, that way, $l_{sf}^{Au:(W,Ta)}$ to be also independent of the temperature for the same reasons that the phonon bath plays only a minor role in the electronic diffusion.

The resonance frequency $f$ \textit{vs.} the magnetic field $H_{res}$ \textcolor{black}{(the derivative resonance spectra acquired for $H_{DC}\approx 100$~mT are reported on Fig.~\ref{FMRSP}a and Fig.~\ref{FMRSPb}a for W and Ta based alloys)} allows to extract the effective magnetic saturation and the in-plane anisotropy in each case. We could observe that the NiFe layer possesses the same effective magnetization compared to the reference Py sample of about $760$ emu.cm$^{-3}$ and a negligible in-plane magnetic anisotropy ($H_{uni}$). Furthermore, the contribution due to inhomogeneities is very small, $<0.1$ mT, and a small but clear increase of the damping constant for Au:W ($4$ nm) can be observed with respect to the reference sample. Thus, we can conclude that the Au:W layer acts as a spin sink layer. This can be verified in the charge current \textit{via} constant $V_C$ measurements from which we have extracted $I_C$, where we may observe a symmetrical Lorentzian current peak at the resonance field \textcolor{black}{(Fig.~\ref{FMRSP}b and Fig.~\ref{FMRSPb}b)}. Such lorentzian dc peak for the current is linearly dependent on the rf-power injected in the cavity. To quantify the spin Hall angle of this material, we used all our experimental values taking advantage of the spin diffusion length already determined and results are reported elsewhere. The bare damping constant for Py was found to be $\alpha_{NiFe}=(6.9 \pm 0.1) ~\times 10^{-3}$ originating from inhomogeneous broadening. This allows us to estimate the effective spin-mixing conductance $g_{eff}$ for Au:W and Au:Ta, respectively and thus the spin-current density $j_s$ injected by SP-FMR. The effective spin mixing conductivity is then extracted to be smaller than the Co/Pt and NiFe/Pt, that is respectively $g_{eff}^{\uparrow\downarrow}\simeq 6\times 10^{18}$~m$^{-2}$ for Au:W and $g_{eff}^{\uparrow\downarrow}\simeq 4\times 10^{18}$~m$^{-2}$ for Au:Ta. A typical factor 5 in the electronic transmission then differs between Co/Pt or NiFe/Pt and NiFe/Au based alloys \textcolor{black}{in favor of Co/Pt}.

However, one must note that in spin-pumping experiments, a smaller value of $g_{eff}^{\uparrow\downarrow}$ does not necessarily implies a much smaller value of the spin-current injected, owing to the larger precession angle of the magnetization for smaller spin-current dissipation \textcolor{black}{at low electronic transparency}. In the present case, a slightly larger spin-Hall charge current $I_C$ extracted  at resonance compared to Co/Pt or NiFe/Pt systems can be understood as the result of a larger spin-Hall angle and larger precession angle despite an effective spin-mixing conductance reduced by more than a factor 3. However, one can anticipate that the THz emission properties are very dependent on the local resistivity $\rho$ of the material, a larger resistivity being detrimental for the observation of an efficient THz signal.

\section{\label{Sec4} Modelling of the THz-TDS spectra.}

\textcolor{black}{The specific exciting pulse laser mode requires some additional comment about modelling. Actually two model pictures are actually admissible based on different physical picture: \textit{i}) a modelling of electron dynamics in the sub picosecond scale, by which the local heating leads to transient precessing of electrons around a local change of the anisotropy magnetic field~\cite{bocklage1,bocklage2}. This constitutes a direct extension to the GHz precession regime, that we do not consider henceforth in that section, or \textit{ii}) a two temperature model by which an electronic absorption gives rise to electronic heating with characteristic excitation energy over the Fermi level (temperature $T_e$ followed by inelastic relaxation due to electron-electron interactions and phonon bath characterized by a second temperature $T_{ph}$~\cite{maldonado2017}. A refined modelling generally requires a \textit{superdiffusive modelling} for excited carriers that we can further consider. Note that in this scenario , the fast energy and spin relaxation rate in the SOC transition metals (Pt, Au:W, Au:Ta) makes that the two different temperature dynamics becomes more and more uncoupled leading to the sole electron dynamics relevant in the THz emission spectra.}

The modelling presented in this section addresses \textit{i}) the time-dependent diffusion and relaxation processes in separate layers of excited spin-polarized carriers, generated by a short laser pulse and characterized by a certain generation rate $\mathcal{P}(\mathbf{r},t)$, \textit{ii}) the reflection/transmission of the spin-polarized hot carriers at the inner interface, as well as \textit{iii}) the specific boundary conditions to consider at the two outward interfaces. Regarding point \textit{ii}), the transmission across the inner interface can involve spin-mixing terms related to spin-flips and spin memory loss like exhibited in FMR spin-pumping experiments.

\subsection{Boltzmann formalism.}

The description of the time-domain dynamics of the hot-electrons within spintronic multilayers excited by ultrashort laser pulses may be performed in the frame of the Boltzmann transport theory and equations (BTE), and considering the different electronic diffusion and relaxation processes. The BTE for spin-polarized particles can be derived as the evolution equation of a reduced single-particle density matrix. It also accounts for the properties of excited electron dynamics in the \textit{sp} bands of very thin metal layers, as widely discussed in the superdiffusive theory of spin currents in both experiments and in modelling~\cite{battiato2010,battiato_theory_2012,maldonado2017,battiato2020}. We start from the evolution equation for the carrier-distribution function in the following form:

\begin{eqnarray}
\left[\frac{\partial}{\partial t}+\frac{\hbar}{m^*}\mathbf{k} \cdot \nabla_\mathbf{r}+\frac{1}{\hbar}\mathbf{E}_\sigma(\mathbf{r},t) \cdot \nabla_\mathbf{k}\right]f_\sigma(\mathbf{r},\mathbf{k},t)=\mathcal{P}_\sigma(\mathbf{r},\mathbf{k},t)\nonumber \\
-\frac{f_\sigma(\mathbf{r},\mathbf{k},t)}{\tau_\sigma(\mathbf{r},E)}+\sum_{\sigma,\sigma^\prime}\int d^3k^\prime ~w(\mathbf{r};k^\prime,\sigma^\prime;\mathbf{k})f_\sigma(\mathbf{r},k^\prime,t) ~ \quad
\label{boltzmann}
\end{eqnarray}
where $f_\sigma$ represents the spin- and time-dependent distribution function in space ($\mathbf{r}$) and in the reciprocal space or Brillouin zone ($\mathbf{k}$), $\mathbf{E}$ the electric field, and where $\mathcal{P}_\sigma (\mathbf{r},\mathbf{k},t)$ is the pump term due to the pulsed laser excitation. In general, the distribution function $f_\sigma=f^0_\sigma+\phi_\sigma$ can be separated into a sum of an equilibrium part $f^0_\sigma(E)$ plus a non-equilibrium (thermal) part $\phi_\sigma(\mathbf{r},\mathbf{k},t)$~\cite{valetfert1993} describing thus the electronic excitations ($\phi_\sigma$ is called g$_\sigma$ in Ref.~\cite{valetfert1993}). The two relaxation terms at the right hand side of Eq.~\eqref{boltzmann} represent the respective \textit{scattering-out} and \textit{scattering-in} processes in the BTE.

We denote the FM layer as the left material (L) while the right one (R) is the heavy metal characterized by a strong SOI, the location of strong spin-relaxation processes. We focus here on a Co/Pt system giving convincing qualitative and quantitative results when compared to the experiments. The ensemble of the physical parameters used are listed in the appendix. The dynamics of the system is performed using the so-called wave-diffusion modelling as described in Refs~\cite{kaltenborn2012,nenno2019} involving the time dynamics of both spin-dependent carrier densities (Eq.~\eqref{evol1}) and spin-currents (Eq.~\eqref{evol2}). We have included relevant relaxation terms and potentially may include local secondary electron sources that we do not consider henceforth. The wave-diffusion model leads to complex differential equations to solve in space and we time considering the same kernel for both spin densities and currents, but admitting a different source term (see Eqs.~\eqref{evol3}).

Eq.~\eqref{evol1} is called the population evolution as it is made of three parts : a spin-flip term associated to inelastic spin-flip scatterings, the current divergence and the spin-dependent source term (laser). This absorption gives rise to excitation of electrons from the $d$ band to unoccupied states from the $sp$ bands above the Fermi level with energy of the order of a fraction of an eV. Besides, Eq.~\eqref{evol2} describes the spin-current dynamics dependent on the momentum relaxation time $\tau_\sigma^{i}$. Moreover, we may include re-magnetization process at a larger timescale $\tau_r$ by adding a supplementary $\textit{sp}\rightarrow \textit{d}$ relaxation term. We thus consider the two-coupled equations in the time-domain, valid for both layers $i=\{\text{L},\text{R}\}$, according to:

\begin{eqnarray}
    \frac{\partial n^i_\sigma}{\partial t}=- \underbrace{\frac{n^i_\sigma - n^i_{\overline{\sigma}}}{\tau^i_{sf}}}_{\text{spin-flip}}
    - \underbrace{\nabla j^i_\sigma}_{\text{flux term}}
    + \left( \underbrace{\mathcal{P}_\sigma (\mathbf{r},t)}_{\text{source term}} - \underbrace{\frac{n^{i}_\sigma}{\tau^{i}_{r}}}_{\text{remag}} \right) \delta _{i,L}  \nonumber \\
    \frac{\partial j^i_\sigma}{\partial t}=-\frac{j^i_\sigma}{\tau_\sigma^i} - \frac{D^i_\sigma}{\tau_\sigma^i} \nabla n^i_\sigma \quad \label{evol2}
\end{eqnarray}
which has to be fulfilled. $\sigma=\left\{\uparrow,\downarrow \right\}$ stands for the spin channel with $\overline{\sigma}=-\sigma$, $n_\sigma= n_ \sigma (\textbf{r},t)$ the population density for each spin channel and $j_\sigma=j_\sigma (\mathbf{r},t)$ the current density for each spin channel. $D_\sigma=\frac{1}{3} v_{F}^2 \tau_\sigma$ is the spin-dependent diffusion coefficient in the corresponding material with $v_F$, the characteristic Fermi velocity of the \textit{sp} band. $\tau_{sf}$ is the spin scattering time and $\mathcal{P}_\sigma (\mathbf{r},t) = \mathcal{P}_\sigma (t)$ the spin-dependent source term that we assume to be homogeneous over the whole ferromagnetic layer thickness.

- In the limit $t<\tau_r$, these coupled equations lead to the determination of both out of equilibrium magnetization (the so-called \textit{spin-accumulation}) $m=n_\uparrow-n_\downarrow$ and spin-current $j_s=j_\uparrow - j_\downarrow$ \textit{via} the following second-order differential equations:

\begin{eqnarray}
    \frac{\partial^2 m}{\partial t^2}+\left[\frac{1}{\tau}+\frac{1}{\tau_{sf}}\right]\frac{\partial m}{\partial t}+\frac{m}{\tau \tau_{sf}}-\frac{\tilde{D}}{\tau}\frac{\partial^2 m}{\partial z^2}=\frac{\partial \mathcal{P}}{\partial t}\\
    \frac{\partial^2 j_s}{\partial t^2}+\left[\frac{1}{\tau}+\frac{1}{\tau_{sf}}\right]\frac{\partial j_s}{\partial t}+\frac{j_s}{\tau \tau_{sf}}-\frac{\tilde{D}}{\tau}\frac{\partial^2 j_s}{\partial z^2}=-\frac{\tilde{D}}{\tau}\frac{\partial \mathcal{P}}{\partial z}
\label{evol3}
\end{eqnarray}
for large $\tau_r$. $\tilde{D}=\frac{2 D_\uparrow D_\downarrow}{D_\uparrow + D_\downarrow}$ is the average diffusion constant, $\tilde{\tau}=\frac{2 \tau_\uparrow \tau_\downarrow}{\tau_\uparrow + \tau_\downarrow}$ the average scattering time and $\mathcal{P}(t)= \mathcal{P}_\uparrow - \mathcal{P}_\downarrow$ the time-profile of the spin-polarized pump. Note that for a steady state regime whereby all quantities no longer depend on time, one obtains the standard space-dependent exponential variation, $m,~j_s\propto \exp\left(\pm \frac{z}{l_{sf}}\right)$ with an exponential decay given by the spin diffusion length, related to the spin-lifetime and diffusion coefficient $\tilde{D}$ according to $l_{sf}=\sqrt{\tilde{D}\tau_{sf}}$.

- \textcolor{black}{The far-field THz electric field $E_{THz}$ emitted is proportional to the time derivative of the local charge current $j_c$ and it is derived from the following expression:}

\begin{eqnarray}
E_{THz} \propto \frac{\partial \int j_c dz}{\partial t}&=& -\theta_{SHE}\int \left( \frac{j_s}{\tau}+\frac{\tilde{D}}{\tau}\frac{\partial m(z,t)}{\partial z}\right) dz
\label{Efield}
\end{eqnarray}
with the charge current $j_c^{3D} \simeq \theta_{SHE} j_s$ proportional to the spin current for ISHE and $\int dz$ means that one operates the integral over the thickness of the HM layer corresponding to a summation over all the emitting dipole. \textcolor{black}{On the other hand, by operating a Fourier transform (FT) on the Eqs.~[\ref{evol2}] acting respectively on $m(z,t)=n_\uparrow(z,t)-n_\downarrow(z,t)=\int \tilde{m}(q,\omega)\exp\left[i(qz-\omega t)\right]dq~d\omega$ and $j_s(z,t)=j_\uparrow(z,t)-j_\downarrow(z,t)=\int \tilde{j}_s(q,\omega)\exp\left[i(qz-\omega t)\right]dq~d\omega$, one may derive:}

\begin{eqnarray}
\tilde{m}(z=0,\omega)&\cong&\frac{\kappa \tau_{sf}}{1-i\omega\tau_{sf}}\tilde{j}_s(z=0,\omega)\\
\tilde{j}_s(z=0,\omega)&=&-\frac{D}{1-i\omega\tau}\nabla_z \tilde{m}(z=0,\omega)
\label{Efield1}
\end{eqnarray}
where $\kappa^{-1}=\left(iq\right)^{-1}\cong l_{sf}=\sqrt{D\tau_{sf}}$ is the typical spin diffusion length over which dipoles are active. From Eq.~[\ref{Efield1}], and by expressing the longitudinal conductivity $\sigma=\sigma_{xx}=e^2D\mathcal{N}$ ($\mathcal{N}$ is the density of states), one may thus derive:

\begin{eqnarray}
E_{THz} &\propto& \frac{i\theta_{SHE}\omega\tau_{sf}D\kappa}{\left(1-i\omega\tau\right)\left(1-i\omega \tau_{sf}\right)} \tilde{j}_s(z=0,\omega)\nonumber\\
E_{THz} &\simeq& \frac{i\omega\tau_{sf}\kappa}{\left(1-i\omega\tau\right)\left(1-i\omega\tau_{sf}\right)} \left(\frac{\sigma_{SHE}}{\mathcal{N}(E)e^2}\right)  \tilde{j}(\omega,z=0)
\label{Efield2}
\end{eqnarray}
\textcolor{black}{Note that for HM materials like Pt, one has $\tau_{sf}\gtrsim \tau$. From Eq.~[\ref{Efield2}], it results that, at equivalent amplitude for the source term $\tilde{m}$, the time domain charge current and subsequent electric field are proportional to the diffusion constant and then to the conductivity of the HM layers. In particular, Eq.~[\ref{Efield2}] proves that, unlike the spin-orbit torque (SOT), no real enhancement of the THz signal is expected if the \textit{intrinsic} spin Hall conductivity $\sigma_{SHE}$ of intrinsic SHE materials, like Pt or Au, is kept fixed or even decreases by alloying despite a possible increase of the effective spin Hall angle $\theta_{SHE}$. This partly explains that poorly conductive materials or alloys like Au:W or Au:Ta provides only modest THz emission compared to Co/Pt or NiFe/Pt.}

\subsection{\label{}Ab-initio calculations of transparencies at Co/Pt interfaces.}

First-principles calculations giving the energy-dependence of the electronic transmission coefficient for the Co/Pt(111) interface were performed within the atomic sphere approximation in the Green's function-based tight-binding linear muffin-tin orbital (GF-LMTO) method \cite{ASA,Turek,Questaal}, treating exchange and correlation within the local-density approximation (LDA) \cite{Barth_1972}. The Green's functions are represented in a mixed basis: the two-dimensional translational periodicity of the interface allows one to introduce the conserved wave vector component $\mathbf{k}_\parallel$ parallel to the interface, which is confined to the two-dimensional Brillouin zone of the interface; the Green's function is then a matrix in real space with arguments confined to the unit cell of the 'active region' including the interface and a few monolayers on each side of it. The active region is embedded between semi-infinite Co and Pt leads. After the self-consistent charge and spin densities were obtained, the total and $\mathbf{k}_\parallel$-resolved energy-dependent ballistic conductance of the Co/Pt bilayer was calculated using the Landauer-B\"uttiker technique.

\begin{figure*}[!htb]
\includegraphics[width=0.9\textwidth]{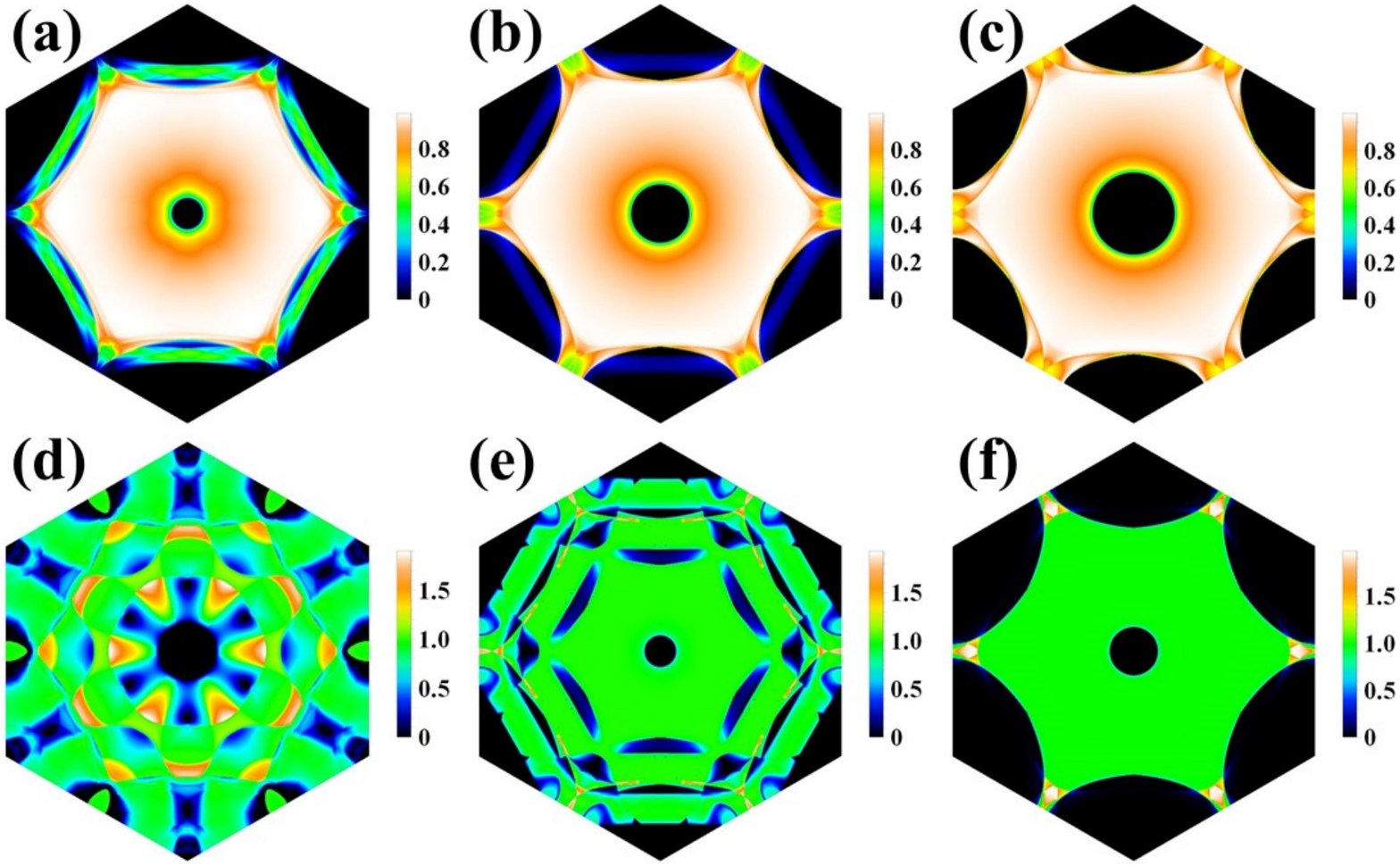}
\caption{Calculated energy-dependent and selected-spin transmission function for the Co/Pt (111) interface plotted as a function of $\mathbf{k}_\parallel$ (projection of the wave vector on the plane of the interface, which is conserved in the scattering process) and summed in the two-dimensional Brillouin zone of the interface. (a-c): Majority-spin, (d-f): minority-spin channel. Panels (a) and (d): $E-E_F=0$, (b) and (e): 0.5 eV, (c) and (f): 1.0 eV. \textcolor{black}{Those calculations shows that near the Fermi level (a-d), the majority electrons (spins $\uparrow$) are more easily transmitted (in average) than the minority spin channel ($\downarrow$)}, in particular near the Brillouin zone center, whereas at higher energy,  0.5 eV above the Fermi level  (e-f), the electronic transmission for the minority spin channel is larger (refer to Fig.~\ref{transm_DOS})}).
\label{k_transm}
\end{figure*}

\begin{figure}[!htb]
\includegraphics[width=0.6\textwidth]{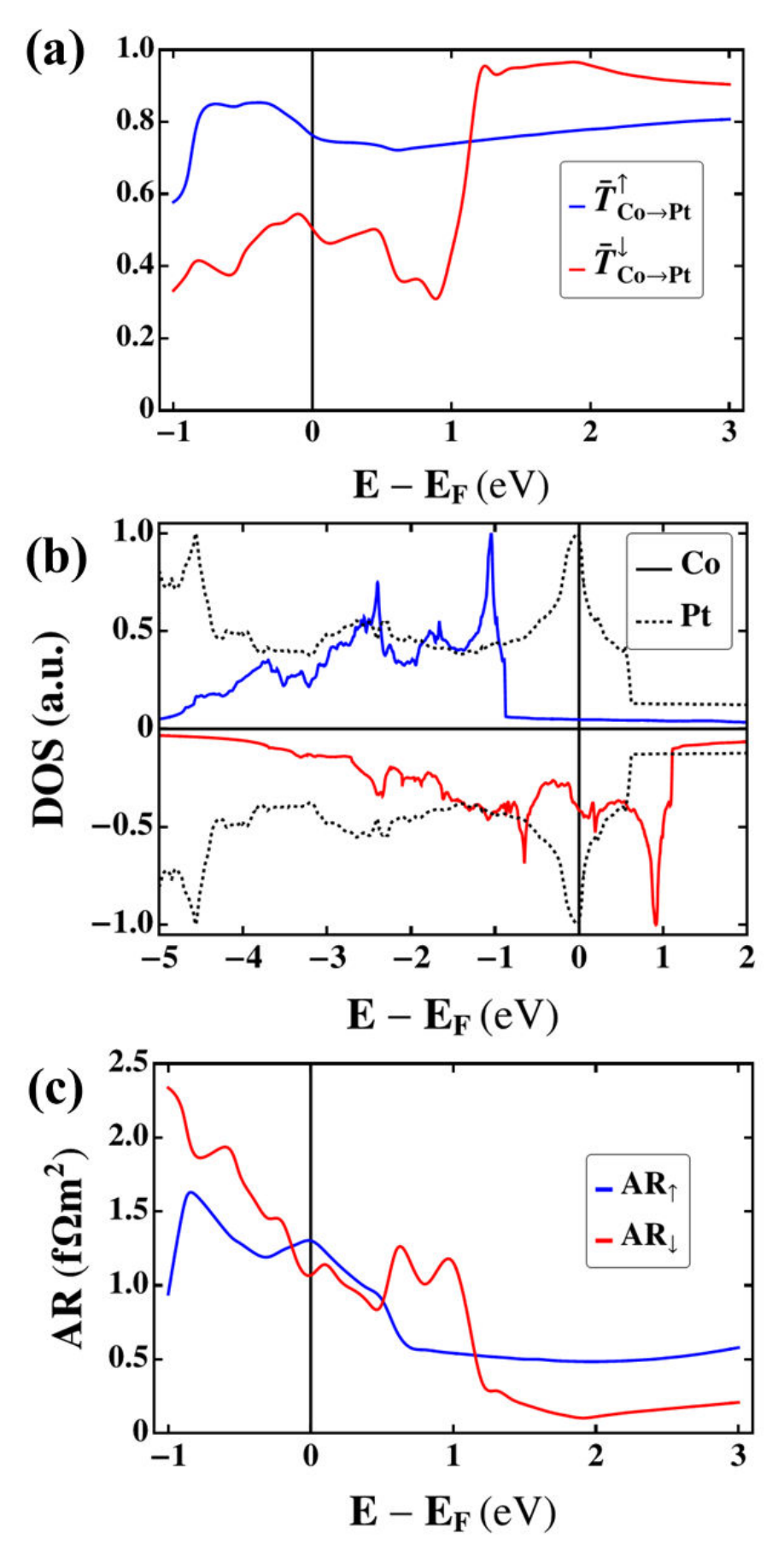}
\caption{\textcolor{black}{(a) Spin and energy-dependent average transmission probability for electrons incident from the Co side on the Co/Pt (111) interface, see Eq. (11). Blue (red) lines: majority (minority) spin electrons. (b) Density of states in bulk fcc Co and Pt with the lattice parameter used in the calculations for the interface. Blue (red) solid lines: majority-spin (minority-spin) electrons in fcc Co. Dotted lines: both spins in Pt. (c) Effective spin- and energy-dependent area-resistance product of the Co/Pt (111) interface. Blue (red) line: majority (minority) spin.}}
\label{transm_DOS}
\end{figure}

We assumed the Co/Pt bilayer has a perfect continuous face-centered cubic lattice with an abrupt (111) interface and a common lattice parameter of $2.64$~\r{A}, ignoring strain relaxation. The average transmission probability for electrons incident from the Co side is:

\textcolor{black}{
\begin{equation}
    \bar{T}_\sigma^{\mathrm{Co}\to\mathrm{Pt}}=\frac{1}{N_\sigma^\mathrm{Co}}\sum_n T_\sigma^n = \frac{g_\sigma}{g_\sigma^{\mathrm{Co}}}
    \label{tbar}
\end{equation}
where $\sigma$ denotes the spin channel, $N_\sigma^\mathrm{Co}$ is the number of conducting channels of a given spin in the Co lead, $T_\sigma^n$ the transmission probability for one of these channels $n$, $g_\sigma$ the conductance of the Co/Pt bilayer, and  $g_\sigma^{\mathrm{Co}}$ the Sharvin conductance of the Co lead. The $\mathbf{k}_\parallel$-resolved transmission function for the Co/Pt (111) interface is shown in Fig.\ \ref{k_transm}, for each spin channel, at 0, 0.5, and 1.0 eV above the Fermi level. One of a relevant quantity for the electronic transport property is the average transmission probability (Eq. \eqref{tbar}) for each spin channel is shown in Fig.\ \ref{transm_DOS}a. The origin of the features observed in these plots can be correlated with the density of states (DOS) of bulk Co and Pt, as discussed below.}

Cobalt has one majority-spin Fermi-surface sheet with topology identical to that of Cu: the somewhat distorted free-electron-like Fermi surface does not quite fit into the first Brillouin zone, forming 'necks' centered around the L points on its hexagonal faces. One of these L points projects onto the center ($\bar\Gamma$ point) of the surface Brillouin zone of the (111) interface, resulting in a circular ``hole'' in the transmission function [panels (a-c) of Fig.~\ref{k_transm}] where there are no bulk states in Co. As the energy increases, the iso-energetic surface expands, and the holes become larger. Panels (a-c) in Fig.~\ref{k_transm} show that most the states from the majority-spin Fermi surface sheet in Co have a high probability of transmission across the interface in the entire energy window shown in Fig.~\ref{transm_DOS}(a). This feature is similar to the well-known case of the  Co/Cu interface \cite{Schep1997} where the band structure match in the majority-spin channel is nearly perfect.

Because Pt has one electron fewer than Cu and open 5\textit{d} shells, its Fermi surface is more complicated, with one free-electron-like electronic sheet, one extended hole-like sheet that projects to the periphery of the (111) surface Brillouin zone, and one small hole-like pocket around the X point. As the energy is increased, the hole-like sheets shrink and disappear at about 0.7 eV where DOS has a van Hove singularity (see Fig.~\ref{transm_DOS}(b)). The majority-spin transmission $\bar T_\uparrow^{\mathrm{Co}\to\mathrm{Pt}}$ is essentially unaffected by the disappearance of the hole-like sheets in Pt, because, as is clear from Fig.~\ref{k_transm}(a-c), the majority-spin transmission is dominated by electrons from Co transmitting into the electron-like sheet in Pt. On the other hand, $\bar T_\downarrow^{\mathrm{Co}\to\mathrm{Pt}}$ decreases significantly in the 0.5-0.7 eV range and remains suppressed up to about 1.0 eV, where it quickly rises up to almost 1. The drop around 0.6 eV is associated with the closure of the hole-like sheet in Pt, which decreases the transmission probability of the minority-spin states from Co. The rise at 1.0-1.2 eV is due to the closure of the similar hole-like minority-spin sheets in Co, the electrons from which transmit poorly into Pt but contribute significantly to the Sharvin conductance of Co.

Fig.~\ref{transm_DOS}(a) shows that the Co/Pt (111) interface has this favorable property in the energy window from 0 to 1.0 eV above the Fermi level, and especially in the 0.6-1.0 eV range, between the top of the 5$d$ band of Pt and the top of the minority-spin $3d$ band of Co. Fig.~\ref{transm_DOS}(c) also shows the spin-resolved effective interfacial area-resistance product calculated according to Schep \textit{et al.} \cite{Schep1997} This quantity represents the apparent spin-dependent resistance of the interface in the circuit under diffusive transport conditions. Although these conditions are not satisfied in THz emission devices, and transport of hot electrons brings its own complications, the spin-dependent effective $RA$ product can serve as an approximate indicator of the interfacial spin asymmetry (\textcolor{black}{noted as the $\gamma$ parameter}). Here we also see that the Co/Pt (111) interface has a larger effective resistance in the minority-spin channel for energies in the 0.6-1.1 eV range that is in the energy region of hot electrons generated by pulsed laser excitations, suggesting \textcolor{black}{an additional 'spin-filtering' effect} and an enhancement of the spin current by this interface. \textcolor{black}{Note however, that the calculation of the spin-selected transmission coefficient per electronic channel (Fig.~\ref{transm_DOS}(a)) also evidence a positive spin-asymmetry coefficient near the Fermi level (majority spin $\uparrow$ are favorably transmitted per channel), and such value of $\gamma\approx +0.3$  found here was recently extracted from refined Anomalous Hall effect (AHE) experiments and subsequent analyses~\cite{dang2020}.}

\section{\label{Sec5} Results of the numerical simulations.}

\subsection{FDTD simulations in the time domain.}

Solving the time-dependent equations requires the implementation of a numerical routine. The experimental boundary conditions corresponding to pulse laser excitation is considered by a zero $\sigma$-spin population within the whole multilayers at $t=0$. Additional treatment of the external and internal boundary conditions is available in Appendix \ref{Boundary}. We have considered a typical temporal Gaussian shape:

\begin{equation}
    \mathcal{P}{^0_\sigma}(\mathbf{r},t) = \mathcal{P}^{0}_{\sigma}(t)  = s_\sigma \frac{A}{\sqrt{2 \pi \Delta t^2 }} \exp{\left[-\frac{(t-4 \Delta t)^2}{2 \Delta t^2} \right]}
    \label{laser}
\end{equation}
uniformly exciting the ferromagnetic material with a wavelength $\lambda \simeq 810$~nm. $A$ is related to the laser pump power, $\Delta t$ the laser pulse duration (typically 100~fs) and $s_\sigma$ the initial proportion of the spin-channel excited owing to the different density of states $\mathcal{N}_{d,\sigma}$ of the \textit{d}-band. \textit{Superdiffusive} spin currents and secondary electron sources can be performed by adding supplementary nonlinear secondary source terms $\mathcal{P}^\prime_\sigma (\mathbf{r},t)$ that we do not consider henceforth.

\begin{table*}[!h]
\centering
\scalebox{0.8}{
\begin{tabular}{lllll}
\textbf{Parameter} & \textbf{Use} & \textbf{Material} & \textbf{Value} & \textbf{Reference} \\
Ferromagnetic layer thickness $t_{FM}$ & Input & Co & 10 nm & - \\
Lifetime of the majority spins $\tau_\uparrow$ & Input & Co & 22 fs & \cite{wieczorek_separation_2015}\\
Lifetime of the minority spins $\tau_\downarrow$ & Input & Co & 7 fs &  \\
Velocity of the majority spins $v_\uparrow$ & Input & Co & 0.6 nm.fs$^{-1}$ &  \cite{wieczorek_separation_2015}\\
Velocity of the minority spins $v_\downarrow$ & Input & Co & 0.6 nm.fs$^{-1}$ &  \\
Diffusion of the majority spins $D_\uparrow$ & $\frac{1}{3} \tau_\uparrow v_\uparrow^2$ & Co & 2.64 nm$^2$.fs$^{-1}$ & - \\
Diffusion of the minority spins $D_\downarrow$ & $\frac{1}{3} \tau_\downarrow v_\downarrow^2$ & Co & 0.84 nm$^2$.fs$^{-1}$ & - \\
Spin asymmetry $\beta$ & Input & Co & 0.46 & \cite{bass_spin-diffusion_2007}\\
Mean free path $\lambda$ & Input & Co & 8.6 nm & \cite{gall_electron_2016}\\
Spin diffusion length $l_{sf}$ & Input & Co & 12 nm & \cite{bass_spin-diffusion_2007}\\
Spin scattering time $\tau_{sf}$ & $\frac{6 l_{sf}^2}{v_F \lambda (1-\beta)^2}$ & Co & 75 fs & - \\
Non magnetic layer thickness $t_{NM}$ & Input & Pt & 4 nm & - \\
Carrier lifetime $\tau= \tau_\uparrow = \tau_\downarrow$ & Input & Pt  & 10 fs & \cite{zhukov2006}\\
Carrier velocity $v=v_\uparrow=v_\downarrow$ & Input & Pt & 0.35 nm.fs$^{-1}$ & \cite{zhukov2006}\\
Diffusion coefficient $D=D_\uparrow=D_\downarrow$ & $\frac{1}{3} \tau v^2$ & Pt & 0.3 nm$^2$.fs$^{-1}$ & - \\
Mean free path $\lambda$ & Input & Pt & 2 nm & \cite{zhukov2006}\\
Spin diffusion length $l_{sf}$ & Input & Pt & 3 nm & \cite{jaffres2014}\\
Spin scattering time $\tau_{sf}$ & $\frac{6 l_{sf}^2}{v_F \lambda}$ & Pt & 20 fs& \cite{zhukov2006}\\
Interfacial spin asymmetry $\gamma$ & Variable & Co/Pt & $+0.5$ & \cite{fert_theory_1996}\\
Interfacial spin resistance $R_B$ & Input & Co/Pt & 0.83 $\times 10^{-15}$ $\Omega$.m$^2$ & \cite{fert_theory_1996}\\
laser amplitude $A$ & Input & - & 1 a.u. & - \\
laser pulse duration $\Delta t$ & Input & - & 100 fs & - \\
Interfacial transmission coefficient $T^\star$ & Variable & Co/Pt & $\left\{ 10^{-4} - 10^{-1}, 0.25, 0.5 \right\}$ & - \\
Transmission of the majority channel $T_\uparrow$ & Deduced & Co/Pt & see Eq. (\ref{T}) & - \\
Transmission of the minority channel $T_\downarrow$ & Deduced & Co/Pt & see Eq. (\ref{T}) & - \\
Remagnetization time $\tau_r$ & Fixed & Co & 10 ps & - \\
Spin-up channel source term $s_\uparrow$ & Variable & Co & $\left\{ 0.5, 1 \right\}$ & - \\
Spin-down channel source term $s_\downarrow$ & Variable & Co & $\left\{ 0.5, 1 \right\}$ & - \\
\end{tabular}}
\caption{Material parameters used in the FDTD simulations.}
\label{para}
\end{table*}

Our numerical investigations will focus on the major role played in the terahertz wave generation by interfacial transmission $T_\sigma$ and characteristic spin-flip times to explain the difference between Co/Pt and NiFe/Au based samples. In a first step, we will describe the shape of the typical THz spectra. For that end, we consider the change in the spectra obtained by varying, in the same way, both the momentum relaxation time $\tau_p$ and the spin-flip time $\tau_{sf}$ from their nominal values $\tau_p^0$ and $\tau_{sf}^0$. We consider the scaling parameter:

\vspace{0.1in}

\begin{equation}
\alpha=\frac{\tau_{sf}}{\tau_{sf}^0}=\frac{\tau_{p}}{\tau_{p}^0}
\end{equation}
in the FM and in the HM material. That permits to simulate the effect of a typical change of the mobility of the different constituents (Pt \textit{vs.} Au-based alloys), keeping fixed their spin-orbit parameter $\epsilon=\frac{\tau_p^0}{\tau_{sf}^0}=\frac{\tau_p}{\tau_{sf}}$. We have previously checked that our simulations give the correct conclusions in the steady-state regime of spin-injection (CW pump).

\begin{figure}[!htb]
\includegraphics[width=0.7\columnwidth]{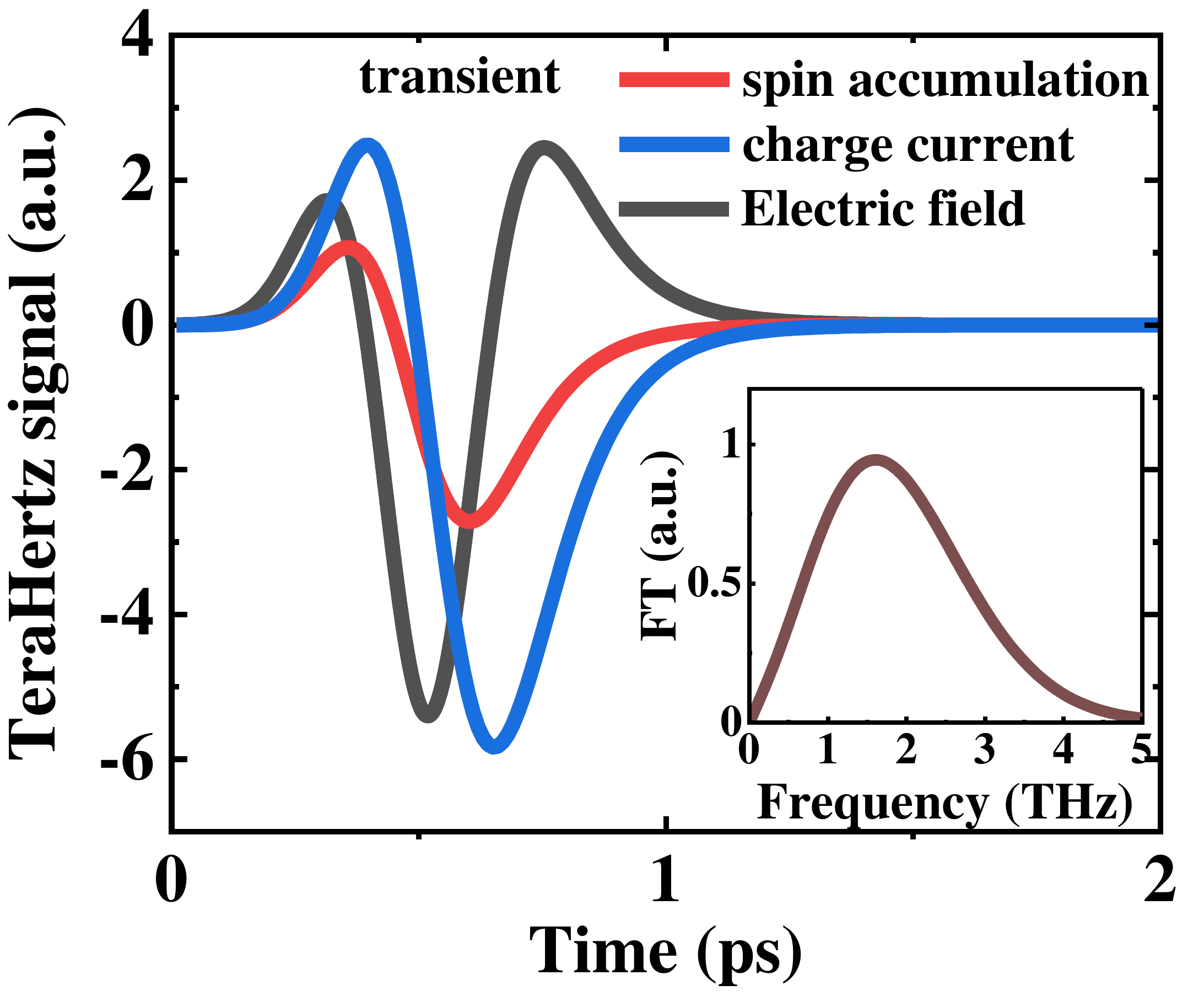}
\caption{\textcolor{black}{Typical simulated time-domain evolution of the emitted electric field terahertz signal (grey) calculated for a Co(2)/Pt(4) structure excited by a 100~fs laser pulse in Co. The generated transient ultrafast surface charge current $j_c$ (blue) and out-of-equilibrium spin accumulation ($m$) (red) are plotted at the interface $z=0+$ on the Pt side. The Fourier transform (FT) of the corresponding terahertz signal is given in the inset. The typical slope of the FT signal at low frequency is representative of a derivative signal ($E_{THz}\propto{\partial J_c}{\partial t}$). In those simulations, inner average transmission coefficient is  $T^\star= 0.2$ and the spin interfacial asymmetry was fixed to $\gamma = 0.5$. The other physical parameters are given in Table I.}}
\label{THZTDS}
\end{figure}

In order to first explore the strong impact of the transmission $T_\sigma$ and the observed differences in the Co/Pt and NiFe/Au:W or NiFe/Au:Ta, we consider the spin average transmission $T^\star$ and spin-asymmetry $\gamma$, as extracted from the ab-initio calculations:

\begin{equation}
\quad T_\sigma = \frac{T^\star}{1 \mp \gamma} \quad \Rightarrow  \quad T^\star = \frac{1-\gamma^2}{2} \left( T_\uparrow + T_\downarrow \right)
\label{T}
\end{equation}
with $T^\star = 0.2$ and $\gamma=+0.5$ at $\epsilon_F$ (Fig.~\ref{transm_DOS}). We first make the assumption that the laser pump generates a ratio of spin $\uparrow$ to spin $\downarrow$ equal to $1:2$ owing to the larger density of states of spin $\downarrow$ electrons at the Fermi level. Fig.~\ref{THZTDS} displays the typical time-dependence of:

\begin{figure*}[!htb]
    \centering
    \includegraphics[width=1\textwidth]{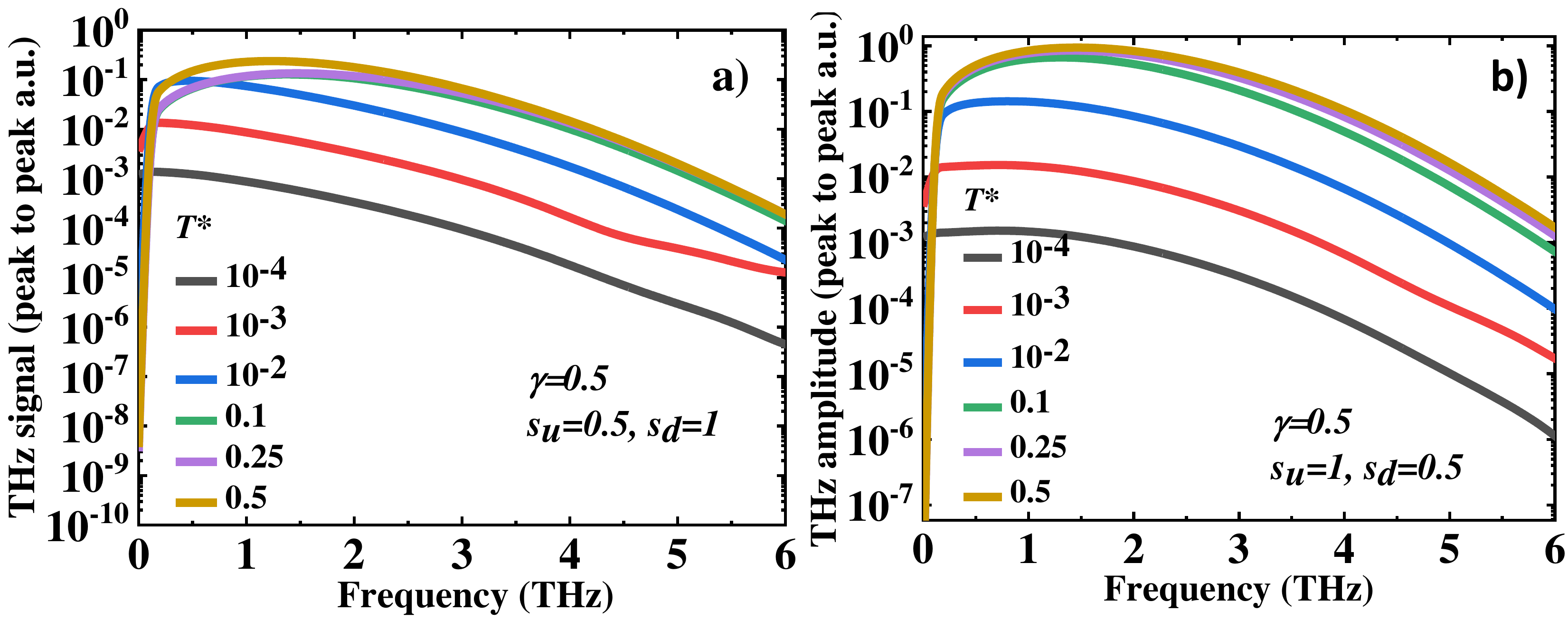}
    \caption{\textcolor{black}{Simulation results showing the frequency-domain representation (Fourier transform: FT) of the emitted terahertz electric field spectra $E_{THz}$ (in log. scale) in the case of Co(2)/Pt(4) excited by a short 100~fs laser pulse and calculated for different average transmission coefficient from $T^*=10^{-4}$ to $T=0.5$. The initial spin polarisation was reversed from (a) $s_\uparrow = 0.5, ~s_\downarrow = 1$ to (b) $s_\uparrow = 1, ~s_\downarrow = 0.5$ (b) while keeping fixed the interfacial spin-asymmetry coefficient $\gamma = +0.5$ through all the simulations. The other physical parameters are garhered on Table I. Those simulations emphasizes on the particular role of the so-caled interfacial 'spin-filtering' effect played by interfaces on the THz spectra.}}
    \label{thztdssimu}
\end{figure*}

- \textit{i}) the out-of-equilibrium spin accumulation or spin-density $m(z=0,t)=n_\uparrow(z=0,t)-n_\downarrow(z=0,t)$ in the HM at the FM/HM interface,

- \textit{ii}) the corresponding ultrafast charge current $j_c(z=0,t)$ in the HM at the FM/HM interface ($z=0$).

 \textit{iii}) $E_{THz}$ in the time domain is also is plotted in the far-field region. This appears to be in close agreement to both \textcolor{black}{Co(2)/Pt(4) and NiFe(2)/Pt(4)} data in shape. The out-of-equilibrium spin-density $m(z=0,t)$ experiences a maximum around $0.4$~ps and a minimum in the vicinity of $t=0.6$~ps. $m(t)$ gives rise to a maximum (minimum) spin current gradually converted into an ultrafast charge current by ISHE, $j_c$. $E_{THz}$ is directly considered as the derivative of the surface charge current $j_c$ integrated within the HM volume. The time response is composed of two small positive lobes surrounding a large negative one resulting from the time-derivative of two Gaussian population functions shifted in time. Regarding the spectral representation of the signal in the frequency domain (see inset), it covers a wide and continuous band up to $5$ THz, and fits reasonably well in form with our experimental data on Co/Pt and NiFe/Pt, taking into account the spectral bandwidth limitation of our detector.

\subsection{FDTD simulations in the frequency domain.}

\emph{Impact of the electronic transmission.}

\vspace{0.1in}

Fig.~\ref{thztdssimu} displays the results of our simulated terahertz spectrum (note the logarithm scale) in the frequency domain using FFT algorithm considering $\gamma = +0.5$. In particular, we have investigated the influence of the inner transmission coefficient $T^\star$ at the FM/HM interface for the two different initial spin polarization of the optical pump, either (a) $s_\uparrow = 0.5, ~s_\downarrow = 1$ (minority spin favorably pump) or (b) $s_\uparrow = 1, ~s_\downarrow = 0.5$ (majority spin favorably pumped). A relevant information is the intensity ratio between subsequent THz spectra \textit{vs.} $T^\star$. For increasing $T^\star$ from $10^{-4}$ to 0.5, the spectra increases in amplitude by about the same proportion. The terahertz signal is the largest for $T^\star = 0.5$ for both pump polarization that is for either minority (Fig.~\ref{thztdssimu}a) and majority pumped spins (Fig.~\ref{thztdssimu}b). Alongside the difference in their conductivity and in their spin-diffusion length, $T^*$ mainly explains the strong difference in the THz spectra between Co/Pt, NiFe/Pt on one side and NiFe:Au:W or NiFe/Au:Ta on the other hand when the transmission for the latter decreases by roughly one order of magnitude (from FMR spin-pumping results). Another important feature is the difference in the signal amplitude (Fig.~\ref{thztdssimu}a \textit{vs.} Fig.~\ref{thztdssimu}b), about one order of magnitude, on reversing the initial polarization and keeping a fixed $\gamma=+0.5$. This feature indicates a preferential spin-selection (filtering) at the interface allowing a larger THz emission when the preferential pumped spin channel and transmitted spin are equivalent.

\begin{figure*}[!htb]
    \centering
    \includegraphics[width=1\linewidth]{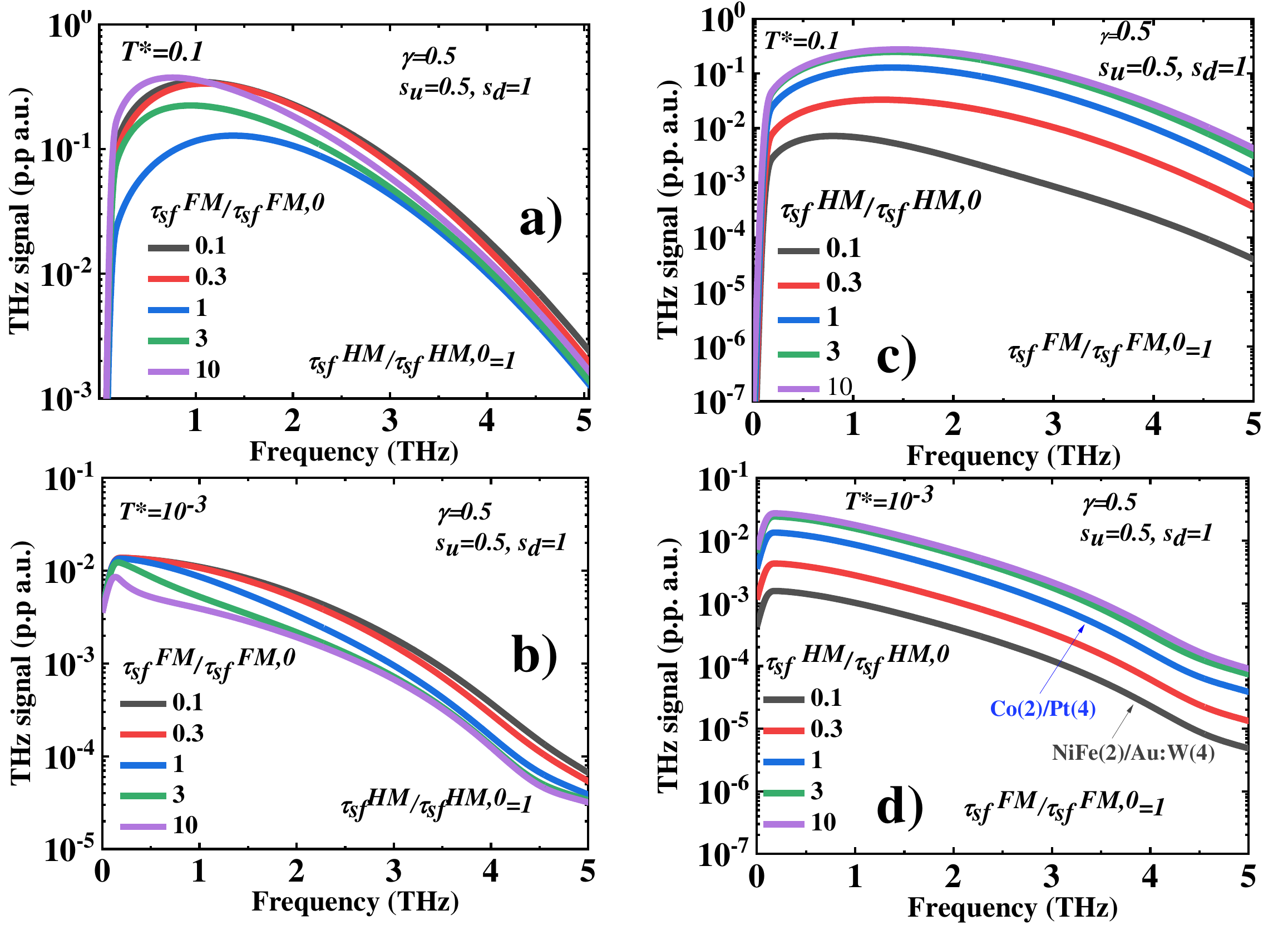}
    \caption{\textcolor{black}{Frequency-domain (FT) representation the emitted electric field ($E_{THz}$) terahertz spectra obtained for Co(2)/Pt(4) after a short 100~fs THz laser pulse and considering an initial spin polarisation $s_\uparrow = 0.5, ~s_\downarrow = 1$. The study focuses on the major role played by the spin dynamics timescale change for a fixed transmission parameter respectively equal tp (i) $T^\star = 10^{-1}$ (a and c) and (ii) $T^\star = 10^{-3}$ (b and d). Those calculations were performed with respect to the transformation Eq.~\ref{transfo} for the ferromagnetic layer characterized by a spin scattering time $\tau_{sf}^{FM}$ (a and b) and for the heavy metal characterized by a spin scattering time $\tau_{sf}^{HM}$ (c and d). The transformation coefficient $\alpha$ is swept in the $[0.1-10]$ range. A constant $\gamma = +0.5$ value was kept for the whole simulations.}}
    \label{THzTDSSimuTau}
\end{figure*}

\vspace{0.1in}

\emph{Effect of material conductivity and spin-flip rate.}

\vspace{0.1in}

On Fig.~\ref{THzTDSSimuTau}, we have explored the effect on the THz spectra of the spin-flip rate in both HM and FM that may be probed in future experiments. One considers here the case of favorably minority spin pumped, $s_\uparrow=0.5$ and $s_\downarrow=1$ and spin-filtering of majority electrons ($\gamma=+0.5$). The typical evolution of THz-TDS spectra resulting from a pump pulse of 100~fs on varying $\alpha = 0.1, 0.3, 1, 3, 10$ in FM and HM are reported on respective Figs.~\ref{THzTDSSimuTau}(a-b)) and Figs.~\ref{THzTDSSimuTau}(c-d)) for $T^\star=10^{-1}$ (Figs.~\ref{THzTDSSimuTau}(a-c)) and $T^\star=10^{-3}$ (Figs.~\ref{THzTDSSimuTau}(b-d)). Two major conclusions can be raised. Concerning the dependence on $\alpha_{FM}$, and for \textcolor{black}{$T^*=0.1$} (Fig.~\ref{THzTDSSimuTau}(a)), one notes a significant increase of the THz signal while increasing $\alpha_{FM}$ from 0.1 to 10 as already pointed out. However, this conclusion is exactly opposite in the case of a small transmission $T^*=10^{-3}$, mainly because the spin-filtering acts in an opposite way to the optical pumping. This yields a reduction of the spin-population in FM and then in the THz spectra (Fig.~\ref{THzTDSSimuTau}(a) \textit{vs.} Fig.~\ref{THzTDSSimuTau}(b)). Second a slight increase of the signal due to spin-flip in FM making the spin-filtering effect more efficient.

On the other hand, increasing the $\alpha_{HM}$ ratio increases the amplitude of the THz spectra (Fig.~\ref{THzTDSSimuTau}(c-d)) owing to a higher mobility, leading to a larger spin-diffusion length and larger volume of charge relaxation in HM. In this sense we demonstrate here the relationship given by Eq.~(\ref{Efield}) by which $E_{THz}$ follows the local HM conductivity (or HM mobility) in proportion to $\alpha_{HM}$. This feature is clearly demonstrated in Fig.~\ref{THzTDSSimuTau}(c-d) whatever $T^\star$ is and also evidenced experimentally when comparing (Co,NiFe)/Pt and NiFe/Au based alloys. Note that the change of $\alpha_{HM}$ does not imply a variation of the spin-orbit parameter ($\epsilon_{HM}$) avoiding thus the possible effect of the so-called impedance mismatch between FM and HM.

\section*{Conclusions}

To conclude, we have demonstrated strong THz pulse emission from Co(2)/Pt(4) and NiFe(2)/Pt(4) transition metal bilayers, mediated by inverse spin-Hall effect in the time domain. Such emission is of dipolar origin and leads to an almost full linear polarization of the electric emitted field. In Pt based interfaces the THz emission is shown to be much larger, by more than one order of magnitude, than that provided by NiFe/Au:W or NiFe/Au:Ta alloys \textit{via} their \textit{extrinsic} ISHE. The difference in the emission efficiency is partly explained from the difference in the resistivity. From our transport simulations, it indeed follows that the optimisation of the THz emission requires a maximum spin-Hall angle (SHA) $\theta_{SHE}$, spin-diffusion length $l_{sf}$ and electronic transmission, $T^\star$, together with a reduced resistivity. It seems that the product $(\theta_{SHE}~\sigma_{xx}~l_{sf}~T^\star)$ constitutes the correct figure of merit for charge conversion in the time-domain. This explains the difference observed between (Co,NiFe)/Pt and NiFe/Au:W or NiFe/Au:Ta based alloys. The above considerations suggest that the performance of the Co/Pt (111) THz emission devices can be increased further by widening the high-asymmetry window between the top of the Pt 5\textit{d} band and the top of the minority-spin Co 3\textit{d} band, while maintaining the large spin-Hall angle in the heavy metal. A natural way to achieve this is to alloy Pt with another element with a higher electron count, such as Au, and to use a ferromagnetic alloy with a larger exchange splitting compared to Co, such as the Fe-Co alloys near the apex of the Slater-Pauling curve. \textcolor{black}{Moreover THz emission spectroscopy methods applied to spintronics structures using inverse spin Hall effect and/or Inverse Edelstein effect may reveal to be a very efficient method to determine the spin injection properties at the nanoscale or in quantum surface states, also to determine the local spin-conductivity within multilayers as performed in standard non-magnetic layers~\cite{hughes2012}.}

\begin{acknowledgements}
We  acknowledge financial support from the Horizon 2020 Framework Programme of the European Commission under FET-Open grant agreement no. 863155 (s-Nebula). T.-H. Dang acknowledges the Horizon2020 Framework Programme of the European Commission under FET-Proactive Grant agreement no. 824123 (SKYTOP). The work at UNL was supported by the National Science Foundation (NSF) through Grant No.\ DMR-1609776, DMR-1916275 and the Nebraska Materials Research Science and Engineering Center (MRSEC, Grant No.\ DMR-1420645). Calculations were performed utilizing the Holland Computing Center of the University of Nebraska, which receives support from the Nebraska Research Initiative.
\end{acknowledgements}

\vspace{0.2in}

\textbf{Data Availability statement}

\vspace{0.1in}

The data that supports the findings of this study are available within the article and its appendices.

{}


\clearpage

\section*{Appendices}

\setcounter{subsection}{0}

\subsection{From continuous pumping to pulse excitation.}

In that appendix, we develop the fundamentals of hot spin-carrier injection in the limit of a continuous pumping (steady state regime of injection) considering characteristic interface transmission and material spin-flip resistances before discussing the case of pulse excitation. We consider \textit{(i)} the spin-injection problem whereby the spin-current is controlled in the FM material far from the interface and \textit{(ii)} the spin-pumping problem corresponding to a constant out-of-equilibrium spin-accumulation at the FM side. The case of a \textit{(iii)} pulsed laser spin-injection will correspond to an intermediate situation combining spin-accumulation and spin-current injections owing its time-domain property.

\subsubsection{Electrical spin-injection}

Concerning the electrical injection issue, the efficiency of spin-injection $\eta_J$ may be defined by the ratio between the spin-current $j^I_s$ injected at the interface (designed as $I$) over its maximum value given in the bulk material $j^\infty_s$. It writes:

\begin{equation}
    \eta_J=\frac{j^I_s}{j^\infty_s}=\frac{g_s r_{FM}}{1+g_s\left(r_{FM}+r_{HM}\right)}\simeq \frac{T^*\sqrt{\frac{\tau_{sf}^{FM}}{\tau^{FM}}}}{{1+T^*\left(\sqrt{\frac{\tau_{sf}^{FM}}{\tau^{FM}}}+\sqrt{\frac{\tau_{sf}^{HM}}{\tau^{HM}}}\right)}}
    \label{spininjection}
\end{equation}
where $g_s=\frac{e^2}{h} \sum_{k_\parallel} T(k_\parallel)$ is the typical surface conductance and $r_s^{FM}$ and $r_s^{HM}$ are the respective spin-resistance (of FM and HM respectively), $r_s=\rho\times l_{sf}\propto \sqrt{\frac{\tau_{sf}}{\tau_p}}$, product of the resistivity by the spin-diffusion length. In the previous expression, $\tau_p$ and $\tau_{sf}$ are the characteristic momentum relaxation time and spin-flip time whereas $\epsilon=\frac{\tau_{p}}{\tau_{sf}}$ can be defined as the spin-orbit parameter (probability of spin-flip after diffusion). The spin resistance represent the resistance to spin-flip. $T^*$ is the average electron transparency over the different incoming channel $k_\parallel$. Eq.~(\ref{spininjection}) involving spin-flip times reflects the proportion of time spent by the carriers in each material separately and weighted by the interface transmission once generated in the FM side. In the limit of the small transparency, the efficiency of spin-injection is found to be small. Moreover, one notes that $\eta_j$ depends strongly on the spin-flip rate in FM, a shorter $\tau_{sf}$ leading to a smaller spin-injection.

\subsubsection{Case of FMR spin-pumping.}

In the case of spin-pumping providing a constant spin-accumulation level in FM of the order of $\Delta \mu \simeq \hbar \omega \sin^2\theta_M$, with $\omega$ is the \textit{rf} frequency and $\theta_M$ the mean precession angle of the magnetization around its equilibrium position, one obtains a spin-injection efficiency $\eta_\mu$ given by:

\begin{equation}
    \eta_\mu=\frac{j^I_s}{\Delta \mu^0}=\frac{g_s}{1+g_s r_s^{HM}}\simeq \frac{T^*}{1+T^*\sqrt{\frac{\tau_{sf}^{HM}}{\tau^{HM}}}}
    \label{spinpumping}
\end{equation}
The efficiency of spin-injection in spin-pumping only depends on the rate of spin-flips in HM and is insensitive to the amount of spin-flip in FM unlike the electrical injection problem. We come to the conclusions that the spin-pumping injection is more robust to material properties, in particular to the properties and spin-flips of the FM material but still dependent of the interface transparency $T^*$.

\subsubsection{Laser pulse injection.}

Because occurring in the time-domain window, the laser pulse injection is controlled by both the spin-current and spin-accumulation. Consequently, the spin-injection efficiency should reveal an intermediate behavior between the electrical and spin-pumping situations. This is what is investigated in the following.

\subsection{time dynamics}

In the limit $t<\tau_r$, these coupled equations lead to the determination of both out of equilibrium magnetization (the so-called \textit{spin-accumulation}) $m=n_\uparrow-n_\downarrow$ and spin-current $j_s=j_\uparrow - j_\downarrow$ \textit{via} the following second-order differential equations:

\begin{eqnarray}
    \frac{\partial^2 m}{\partial t^2}+\left[\frac{1}{\tau_{sf}} +\frac{\delta_{i,L}}{\tau_r} \right] \frac{\partial m}{\partial t} - \frac{1}{\tilde{\tau}} \frac{\partial j_s}{\partial z}-\frac{\tilde{D}}{\tilde{\tau}}\frac{\partial^2 m}{\partial z^2}=\frac{\partial \mathcal{P}}{\partial t} \delta_{i,L} \qquad \nonumber\\
    \frac{\partial^2 j_s}{\partial t^2}+\frac{1}{\tilde{\tau}} \frac{\partial j_s}{\partial t}-\frac{\tilde{D}}{\tilde{\tau}} \left[ \frac{1}{\tau_{sf}} + \frac{\delta_{i,L}}{\tau_r} \right] \frac{\partial m}{\partial z}-\frac{\tilde{D}}{\tilde{\tau}} \frac{\partial^2 j_s}{\partial z^2}=-\frac{\tilde{D}}{\tilde{\tau}}\frac{\partial \mathcal{P}}{\partial z} \delta_{i,L} \qquad\nonumber
\label{evol3}
\end{eqnarray}
or equivalently:

\begin{eqnarray}
    \frac{\partial^2 m}{\partial t^2}+\left[\frac{1}{\tau}+\frac{1}{\tau_{sf}}\right]\frac{\partial m}{\partial t}+\frac{m}{\tau \tau_{sf}}-\frac{\tilde{D}}{\tau}\frac{\partial^2 m}{\partial z^2}=\frac{\partial \mathcal{P}}{\partial t}\\
    \frac{\partial^2 j_s}{\partial t^2}+\left[\frac{1}{\tau}+\frac{1}{\tau_{sf}}\right]\frac{\partial j_s}{\partial t}+\frac{j_s}{\tau \tau_{sf}}-\frac{\tilde{D}}{\tau}\frac{\partial^2 j_s}{\partial z^2}=-\frac{\tilde{D}}{\tau}\frac{\partial \mathcal{P}}{\partial z}
\label{evol3}
\end{eqnarray}
for large $\tau_r$. $\tilde{D}=\frac{2 D_\uparrow D_\downarrow}{D_\uparrow + D_\downarrow}$ is the average diffusion constant, $\tilde{\tau}=\frac{2 \tau_\uparrow \tau_\downarrow}{\tau_\uparrow + \tau_\downarrow}$ the average scattering time and $\mathcal{P}= \mathcal{P}_\uparrow - \mathcal{P}_\downarrow$ the time-profile of the spin-polarized pump.

\subsection{Boundary conditions for spin-polarized hot carriers} \label{Boundary}

In order to establish the dynamics of spin-currents within the bilayers, we must introduce the specific boundary conditions. These consist in the continuity of the spin current $j_s$ at the inner bilayer interface ($z=0$ by definition) at any time $t$ once spin-flips and  spin decoherence have been neglected. For a given spin channel $\sigma$, the related current density at both sides are related to the spin population $n_\sigma$ in both regions and to their characteristic transfer velocity $v^I_{\sigma \sigma^\prime}$ by:

\begin{equation}
\begin{split}
j_{\sigma}^L(z=0^-,t) &= v^I_{\sigma\sigma} \left[n^L_\sigma (z=0^-,t)-n^R_\sigma(z=0^+,t)\right] + v^I_{\uparrow\downarrow} \left[n^L_\sigma (z=0^-,t)-n^R_{\overline{\sigma}}(z=0^+,t)\right] \\
j_{\sigma}^R(z=0^+,t) &= v^I_{\sigma\sigma} \left[n^L_\sigma (z=0^-,t)-n^R_{\sigma}(z=0^+,t)\right]+ v^I_{\uparrow\downarrow} \left[n^L_{\overline{\sigma}} (z=0^-,t)-n^R_\sigma (z=0^+,t)\right]
\label{boundary1}
\end{split}
\end{equation}
where the transfer velocity $v^I_{\sigma \sigma\prime}$ writes $v^I_{\sigma \sigma\prime}=v_F T^{\sigma \sigma^\prime}$ linked to the interface conductance $g^I_{\sigma \sigma^\prime}$ and density of states $\mathcal{N}_{E}$ by $g^I_{\sigma \sigma^\prime}=\frac{e^2}{2\pi} \mathcal{N}_{E}v^I_{\sigma\sigma^\prime}$. In that sense, $\frac{\sqrt{2} T^{\uparrow \downarrow}}{\sqrt{T^{\uparrow \uparrow}+T^{\downarrow \downarrow}}}$ may be viewed as the probability of spin-flip $p_{sf}$. From Eq.~\ref{boundary1}, we note that $j_{\uparrow}^L+j_{\downarrow}^L=j_{\uparrow}^R+j_{\downarrow}^R$ manifests the current continuity at the inner interface \textit{I} whereas $j_s^L-j_s^R=\left(j_{\uparrow}^L-j_{\downarrow}^L\right)-\left(j_{\uparrow}^R+j_{\downarrow}^R\right)=2v_{\uparrow \downarrow}\left(m^L+m^R\right)$ is the condition for the spin-current discontinuity with $m^{L,R}=n_\uparrow^{L,R}-n_\downarrow^{L,R}$ the spin accumulation at both side. We consider pure specular reflections (i.e. $R=1$) on the two external surfaces located at $z=z_B^{\pm}$ (left $-$ and right $+$ limits). This leads to:

\begin{eqnarray}
n_\sigma ^{\text{out}} (z_B^{\pm},t) &=& \left( 1-\frac{p_{sf}}{2} \right) n_\sigma^{\text{in}} + \frac{p_{sf}}{2} n_{\bar{\sigma}} ^{\text{in}} \nonumber\\
j_c(z=z_B^{\pm},t) &=& j_{\uparrow} (z=z_B^{\pm},t) + j_{\downarrow} (z=z_B^{\pm},t) = 0\nonumber \\
j_s(z=z_B^{\pm},t) &=& j_{\uparrow} (z=z_B^{\pm},t) - j_{\downarrow} (z=z_B^{\pm},t) = \frac{p_{sf}}{2-p_{sf}}  \times \left[m(z=z_B^{\pm},t)\right] \left(\frac{v_{F,\uparrow}+v_{F,\downarrow}}{2}\right) ~
\end{eqnarray}
considering $p_{sf}$ as the probability of spin-flip after scattering on the outer surfaces. Note that inside the non-magnetic layer, the carrier velocity is spin-independent.

\subsection{Modelling of spin-transport in the steady-state (continuous pumping).}

\begin{figure*}[!htb]
\includegraphics[width=1\linewidth]{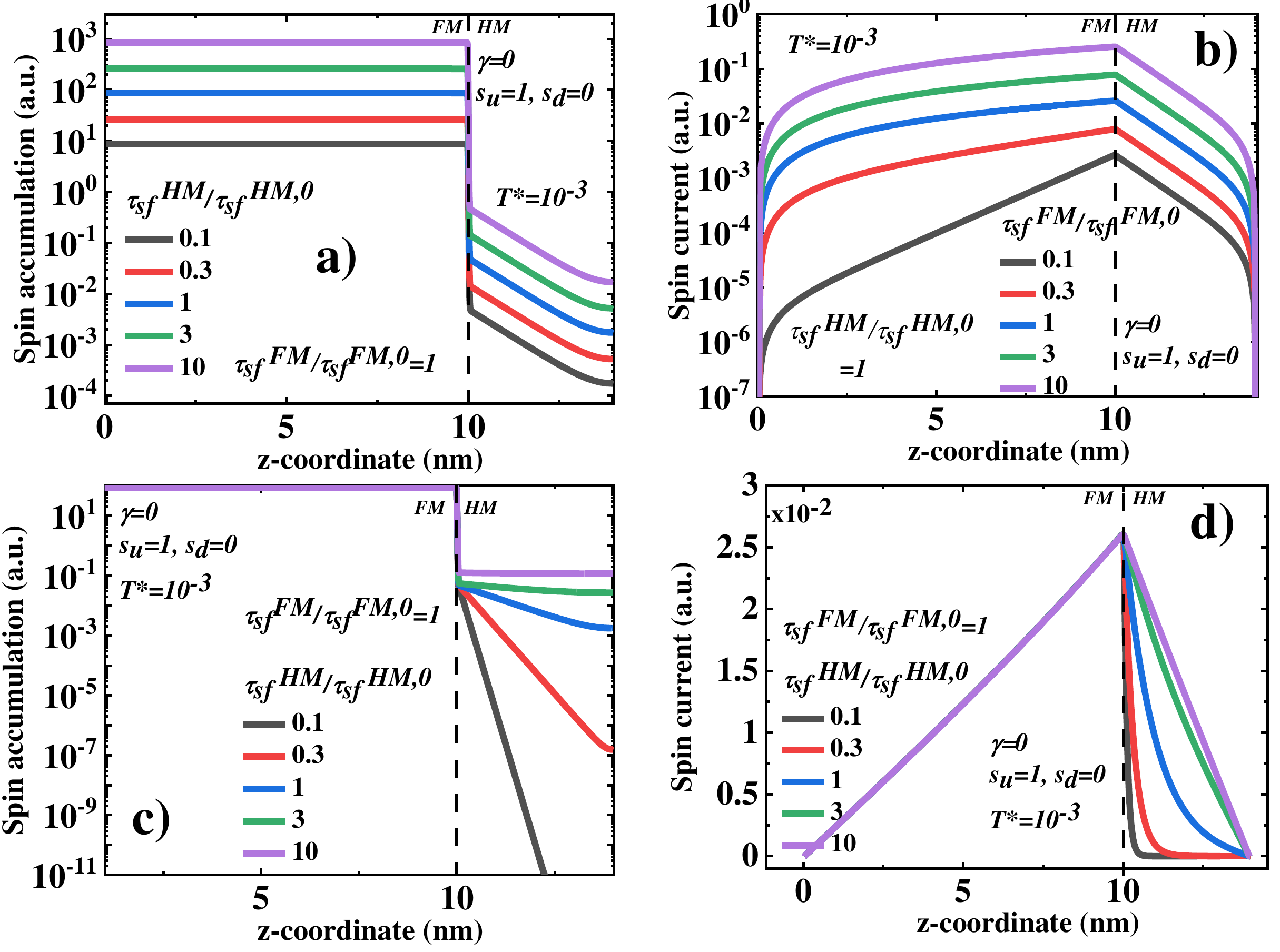}
\caption{Confined case simulation considering a ferromagnetic continuous pumping with $s_\uparrow = 1$ and $s_\downarrow=0$ on a Co(10)/Pt(4) modelled sample. Spin accumulations $m$ (subfigures a and c) and spin currents $j_s$ (subfigures b and d) are presented for long-time approximation by respectively varying $\tau_{sf}^{HM}$ (subfigures a and b) $\tau_{sf}^{FM}$ (subfigures c and d). The interface between the heavy metal and the ferromagnetic layer is modelled using a spin interfacial asymmetry $\gamma=0$ and a coefficient transmission $T^\star = 10^{-3}$.}
\label{THZTDSconfine}
\end{figure*}

In order to analyze our results, we consider the following homothetic time transformation:

\begin{equation}
    \tau_\sigma ' \rightarrow \alpha \tau_\sigma \qquad \text{and} \qquad \tau_{sf}  ' \rightarrow \alpha \tau_{sf}
    \label{transfo}
\end{equation}
scaled by a unitless $\alpha$ parameter defined in each region (FM/HM) separately. We consider here the case of a continuous laser pump in FM in the particular case of a low-transmission $T$ limit at the FM/HM interface. We present on Fig.~\ref{THZTDSconfine} the calculated spin accumulation (or spin-density) profile along $z$, $m(z)$ (Fig.~\ref{THZTDSconfine}(a-c)) alongside the spin current profile $j_s(z)$ (Fig.~\ref{THZTDSconfine}(b-d)) after a long delay time (30~ps). The electronic parameters for both materials (FM/HM=Co/Pt) are gathered in Table~\ref{para}. The sample is typically Co(10)/Pt(4). We have lowered the transmission coefficient $T^\star$ down to $10^{-3}$and considered a null interface transmission spin-asymmetry $\gamma=0$ to simplify. The laser pump is continuously on (i.e. $\mathcal{P}_\sigma(t)=\mathcal{P}=A s_\sigma$) with the result that only hot spin-up population is generated ($s_\uparrow=1$ and $s_\downarrow=0$). In this simulation, we emphasize on the respective role of $\tau_{sf}^{FM}$ (Fig.~\ref{THZTDSconfine}(a-b)) and  $\tau_{sf}^{HM}$ (Fig.~\ref{THZTDSconfine}(c-d)).

One notes several interesting features making the problem particularly singular. First, if one varies $\alpha_{FM}=\frac{\tau_{sf}^{FM}}{\tau_{sf}^{0,FM}}$ by two orders of magnitudes from 0.1 to 10, one notes a strong change in the level of the spin population saturating to a constant value inside the FM (Fig.~\ref{THZTDSconfine}(a)) . The value at saturation of the spin-density population $\Delta \mu\propto m=n_\uparrow-n_\downarrow$ in FM increases quasi-linearly with $\alpha_{FM}$ owing to the correspondingly increase of the spin-relaxation time. Thus FM plays the role of a cavity or spin-reservoir with a particle level fixed by their characteristic spin-flip time. Correspondingly, in HM, the spin population decreases typically by a factor $10^{-3}$. The spin accumulation in HM decreases in proportion to that in FM and to the electronic transmission $T^*$. The spatial decrease of the spin-population profile in HM remains insensitive to the $\alpha_{FM}$ parameter because the spin-diffusion length in HM remains fixed. The spatial dependence of spin-population and spin-current remains unaffected in HM. In the limit of small electron transparency, the spin current space profile $j_s(z)$ calculated by considering the continuity of the spin current at the inner interface and a zero spin current at the edges of the materials follow the same typical decrease than the spin-density (Fig.~\ref{THZTDSconfine}(b)).

\vspace{0.1in}

Fig.~\ref{THZTDSconfine}(c-d)) display the equivalent space profiles when $\alpha_{HM}=\frac{\tau_{sf}^{HM}}{\tau_{sf}^{0,HM}}$ is changed in the same way, from 0.1 to 10, keeping fixed the parameters in the FM layer and still for a small electronic transparency $T=10^{-3}$. Profiles are taken at a time $t=30~ps$. The differences are noticeable from the previous case. The level of spin population in FM remains now constant due to the fixed value $\tau_{sf}^{FM}=\tau_{sf}^{FM,0}$. The spin-current in FM admits a constant profile characterized by a same value at the two FM interfaces ($j_s=0$ to the left and $j_s\approx 2.6 \text{~(a.u)}$ at the FM/HM interface). The main changes with varying $\alpha_{HM}$ are now related to the space-varying profile of both spin-accumulation $m(z)$ and spin-current $j_s(z)$ in $HM$ due to the conjugate variation of $\tau_{sf}^{HM}$ and $\tau_{HM}$. Those space-profile lengthscale is now governed by the effective spin diffusion length $l_{sf}^{HM}=\sqrt{D_{HM}\tau_{sf}^{HM}}\propto \alpha_{HM}$ as expected before saturating to a value close to the thickness $t_{HM}$ owing to the boundary condition chosen ($j_s=0$ at the outward boundaries in particular at the right boundary of Pt).

\vspace{0.2in}


\clearpage

\clearpage

\nocite{*}
\begin{thebibliography}{}

\bibitem{Dhillon_2017} S.~S. Dhillon, M.~S. Vitiello, E.~H. Linfield, and A.~G.~D. et al., Journal of Physics D: Applied Physics \textbf{50}, 043001 (2017).

\bibitem{chen2020} S.~C. Chen, Z. Feng, J. Li, W. Tan, L.-H. Du, J. Cai, Y. Ma, K. He, H. Ding, Z.~H. Zhai \textit{et al.}, Light: Science and Applications \textbf{9}, 99 (2020).
    
\bibitem{Hilton04} D.~J. Hilton, R. D. Averitt, C. A. Meserole, G. L. Fisher, D. J. Funk, J. D. Thompson, and A. J. Taylor, Opt. Lett. \textbf{29}, 1805 (2004).

\bibitem{Seifert_2016} T. Seifert, S. Jaiswal, U. Martens, J. Hannegan, L. Braun, P. Maldonado, F. Freimuth, A. Kronenberg, J. Henrizi, I. Radu, et al., Nature Photonics \textbf{10}, 483–488 (2016).
    
\bibitem{Beaurepaire1996}  E. Beaurepaire, J.~C. Merle, A. Daunois, and J.-Y. Bigot, Phys. Rev. Lett. \textbf{76}, 4250 (1996).

\bibitem{Koopmans2000}  B. Koopmans, M. van Kampen, J. T. Kohlhepp, andW. J. M. de Jonge, Phys. Rev. Lett. \textbf{85}, 844 (2000).

\bibitem{Beaurepaire2004}  E. Beaurepaire, G. M. Turner, S. M. Harrel, M. C. Beard, J.-Y. Bigot, and C. A. Schmuttenmaer, Appl. Phys. Lett. \textbf{84}, 3465 (2004).
    
\bibitem{melnikov2011}  A. Melnikov, I. Razdolski, T.~O. Wehling, E.~T. Papaioannou, V. Roddatis, P. Fumagalli, O. Aktsipetrov, A.~I. Lichtenstein, and U. Bovensiepen, Phys. Rev. Lett. \textbf{107}, 076601 (2011).
    
\bibitem{Huisman2015b} T.~J. Huisman, R.~V. Mikhaylovskiy, A. Tsukamoto, T. Rasing, and A.~V. Kimel, Phys. Rev. B \textbf{92}, 104419 (2015).

\bibitem{Jin2015} Z. Jin, A. Tkach, F. Casper, V. Spetter, H. Grimm, A. Thomas, T. Kampfrath, M. Bonn, M. Kläui, and D. Turchinovich, Nat. Phys. \textbf{11}, 761 (2015).

\bibitem{Huisman2016} T.~J. Huisman, R. V. Mikhaylovskiy, J. D. Costa, F. Freimuth, E. Paz, J. Ventura, P. P. Freitas, S. Blugel, Y. Mokrousov, T. Rasing \textit{et al.}, Nat. Nano. \textbf{11}, 455 (2016).

\bibitem{Yang2016} D. Yang, J. Liang, C. Zhou, L. Sun, R. Zheng, S. Luo, Y. Wu, and J. Qi, Advanced Optical Materials \textbf{4}, 1944 (2016).

\bibitem{saitoh2006} E. Saitoh, M. Ueda, H. Miyajima, and G. Tatara, App. Phys. Lett. \textbf{88}, 182509 (2006),

\bibitem{Hoffmann2013SpinHE} A. Hoffmann, IEEE Transactions on Magnetics \textbf{49}, 5172 (2013).

\bibitem{Sinova2015} J. Sinova, S.~O. Valenzuela, J. Wunderlich, C.~H. Back, and T. Jungwirth, Rev. Mod. Phys. \textbf{87}, 1213 (2015).

\bibitem{Guo2008} G.~Y. Guo, S. Murakami, T.-W. Chen, and N. Nagaosa, Phys. Rev. Lett. \textbf{100}, 096401 (2008).

\bibitem{Tanaka2008} T. Tanaka, H. Kontani, M. Naito, T. Naito, D.~S. Hirashima, K. Yamada, and J. Inoue, Phys. Rev. \textbf{B77}, 165117 (2008).

\bibitem{EDELSTEIN1990233} V. Edelstein, Solid State Communications \textbf{73}, 233 (1990),

\bibitem{Sanchez2013} J. C. R. Sánchez, L. Vila, G. Desfonds, S. Gambarelli, J.~P. Attané, J.~M. De Teresa, C. Magén, and A. Fert, Nat. Comm. 4, 2944 (2013).

\bibitem{Zhou2018} C. Zhou, Y.~P. Liu, Z. Wang, S.~J. Ma, M.~W. Jia, R.~Q. Wu, L. Zhou, W. Zhang, M.~K. Liu, Y. Z.Wu \textit{et al.}, Phys. Rev. Lett. \textbf{121}, 086801 (2018).

\bibitem{Hoffmann2018} M.~B. Jungfleisch, Q. Zhang, W. Zhang, J.~E. Pearson, R.~D. Schaller, H. Wen, and A. Hoffmann, Phys. Rev. Lett. \textbf{120}, 207207 (2018).
    
\bibitem{Hyunsoo2018} X. Wang, L. Cheng, D. Zhu, Y. Wu, M. Chen, Y. Wang, D. Zhao, C.~B. Boothroyd, Y. M. Lam, J.~X. Zhu \textit{et al.}, Advanced Materials \textbf{30}, 1802356 (2018).
    
\bibitem{Hibberd2019} M.~T. Hibberd, D.~S. Lake, N.~A.~B. Johansson, T. Thomson, S.~P. Jamison, and D.~M. Graham, Appl. Phys. Lett. \textbf{114}, 031101 (2019).

\bibitem{wang2019} B. Wang, S. Shan, X. Wu, C.Wang, C. Pandey, T. Nie, W. Zhao, Y. Li, J. Miao, and L.Wang, Appl. Phys. Lett. \textbf{115}, 121104 (2019).

\bibitem{Hyunsoo2017} Y. Wu, M. Elyasi, X. Qiu, M. Chen, Y. Liu, L. Ke, and H. Yang, Adv. Mat. \textbf{29}, 1603031 (2017).

\bibitem{Seifert2017} T. Seifert, S. Jaiswal, M. Sajadi, G. Jakob, S. Winnerl, M. Wolf, M. Klaui, and T. Kampfrath, Appl. Phys. Lett. \textbf{110}, 252402 (2017).

\bibitem{Seifert_2018} T.~S. Seifert, N.~M. Tran, O. Gueckstock, S. M.~Rouzegar, L. Nadvornik, S. Jaiswal, G. Jakob, V.~V. Temnov, M. Munzenberg, M. Wolf \textit{et al.}, Journal of Physics D: Applied Physics \textbf{51}, 364003 (2018).

\bibitem{Torosyan2018} G. Torosyan, S. Keller, L. Scheuer, R. Beigang, and E.~T. Papaioannou, Sci. Rep. \textbf{8}, 1311 (2018).

\bibitem{Qiu18} H.~S. Qiu, K. Kato, K. Hirota, N. Sarukura, M. Yoshimura, and M. Nakajima, Opt. Express \textbf{26}, 15247 (2018).

\bibitem{Cramer2018} J. Cramer, T. Seifert, A. Kronenberg, F. Fuhrmann, G. Jakob, M. Jourdan, T. Kampfrath, and M. Kläui, Nano Letters \textbf{18}, 1064 (2018).

\bibitem{Herapath2019} R.~I. Herapath, S. M. Hornett, T. S. Seifert, G. Jakob, M. Kläui, J. Bertolotti, T. Kampfrath, and E. Hendry, Appl. Phys. Lett. \textbf{114}, 041107 (2019).

\bibitem{hyunsoo2018b} M. Chen, Y. Wu, Y. Liu, K. Lee, X. Qiu, P. He, J. Yu, and H. Yang, Advanced Optical Materials \textbf{7}, 1801608 (2019).

\bibitem{feng2018} Z. Feng, R. Yu, Y. Zhou, H. Lu,W. Tan, H. Deng, Q. Liu, Z. Zhai, L. Zhu, J. Cai \textit{et al.}, Advanced Optical Materials \textbf{6}, 1800965 (2018).
    
\bibitem{cornell2019c} L. Zhu, D.~C. Ralph, and R.~A. Buhrman, Phys. Rev. \textbf{B99}, 180404 (2019).

\bibitem{cornell2019b} L. Zhu, L. Zhu, S. Shi, M. Sui, D. Ralph, and R. Buhrman, Phys. Rev. Applied \textbf{11}, 061004 (2019).

\bibitem{cornell2019a} L. Zhu and R. Buhrman, Phys. Rev. Applied \textbf{12}, 051002 (2019).

\bibitem{Lee2019} H.~Y. Lee, S. Kim, J.~Y. Park, Y.~W. Oh, S.~Y. Park, W. Ham, Y. Kotani, T. Nakamura, M. Suzuki, T. Ono \textit{et al.}, APL Materials \textbf{7}, 031110 (2019).
    
\bibitem{zhucornell2019} L. Zhu, D. C. Ralph, and R. A. Buhrman, Phys. Rev. Lett. \textbf{122}, 077201 (2019).

\bibitem{jaffres2014} J.~C. Rojas-Sánchez, N. Reyren, P. Laczkowski, W. Savero, J.~P. Attané, C. Deranlot, M. Jamet, J.~M. George, L. Vila, and H. Jaffrès, Phys. Rev. Lett. \textbf{112}, 106602 (2014).

\bibitem{berger2018} A.~J. Berger, E.~R.~J. Edwards, H.~T. Nembach, O. Karis, M. Weiler, and T.~J. Silva, Phys. Rev. \textbf{B98}, 024402 (2018).

\bibitem{belashchenko2016} K.~D. Belashchenko, A.~A. Kovalev, and M.~van Schilfgaarde, Phys. Rev. Lett. \textbf{117}, 207204 (2016).

\bibitem{nikolic2018} K. Dolui and B.~K. Nikolic, Phys. Rev. \textbf{B96}, 220403 (2017).

\bibitem{Zeng2019} M. Zeng, B. Chen, and S.~T. Lim, Appl. Phys. Lett. \textbf{114}, 012401 (2019).

\bibitem{kelly2019} R.~J.~H. Wesselink, K. Gupta, Z. Yuan, and P.~J. Kelly, Phys. Rev. \textbf{B99}, 144409 (2019).

\bibitem{kelly2020} K. Gupta, R.~J.~H. Wesselink, R. Liu, Z. Yuan, and P.~J. Kelly, Phys. Rev. Lett. \textbf{124}, 087702 (2020).

\bibitem{Liu2011} L. Liu, T. Moriyama, D.~C. Ralph, and R.~A. Buhrman, Phys. Rev. Lett. \textbf{106}, 036601 (2011).

\bibitem{parkin2015} W. Zhang, W. Han, X. Jiang, S.-H. Yang, and S.~S.~P. Parkin, Nat. Phys. \textbf{29}, 496–502 (2015).

\bibitem{ohno2019} A. Okada, Y. Takeuchi, K. Furuya, C. Zhang, H. Sato, S. Fukami, and H. Ohno, Phys. Rev. Applied \textbf{12}, 014040 (2019).

\bibitem{kamp2020} A. Ciuciulkaite, O. Gueckstock, A. Ravensburg, M. Pohlit, T. Warnatz, T. Kampfrath, G. Schmidt, E.~T. Papaioannou, and V. Kapaklis, Arxiv:2010.12457v1 (2020).

\bibitem{nenno2019} D.~M. Nenno, L. Scheuer, D. Sokoluk, S. Keller, G. Torosyan, A. Brodyanski, J. Lösch, M. Battiato, M. Rahm, R.~H. Binder, et al., Sci. Rep. \textbf{9}, 13348 (2019).

\bibitem{battiato2020} W.-T. Lu, Y. Zhao, M. Battiato, Y. Wu, and Z. Yuan, Phys. Rev. \textbf{B101}, 014435 (2020).

\bibitem{lesne2016} E. Lesne, Y. Fu, S. Oyarzun, J. C. Rojas-Sanchez, D.~C. Vaz, H. Naganuma, G. Sicoli, J.~P. Attané, M. Jamet, E. Jacquet \textit{et al.}, Nat. Mater. \textbf{15}, 1261 (2016).
    
\bibitem{rojasapl} J.~C. Rojas-Sánchez, P. Laczkowski, J. Sampaio, S. Collin, K. Bouzehouane, N. Reyren, H. Jaffrès, A. Mougin, and J.~M. George, Appl. Phys. Lett. \textbf{108}, 082406 (2016).

\bibitem{laczkowski2014}  P. Laczkowski, J.~C. Rojas-Sánchez, W. Savero-Torres, H. Jaffrès, N. Reyren, C. Deranlot, L. Notin, C. Beigné, A. Marty, J.~P. Attané, et al., Appl. Phys. Lett. \textbf{104}, 142403 (2014).

\bibitem{laczkowski2015} P. Laczkowski, H. Jaffrès, W. Savero-Torres, J.-C. Rojas-Sánchez, Y. Fu, N. Reyren, C. Deranlot, L. Notin, C. Beigné, J.-P. Attané \textit{et al.}, Phys. Rev. \textbf{B92}, 214405 (2015).

\bibitem{laczkowski2017} P. Laczkowski, Y. Fu, H. Yang, J.-C. Rojas-Sánchez, P. Noel, V. T. Pham, G. Zahnd, C. Deranlot, S. Collin, C. Bouard \textit{ et al.}, Phys. Rev. \textbf{B96}, 140405 (2017).
    
\bibitem{back2016} M. Obstbaum, M. Decker, A. K. Greitner, M. Haertinger, T.~N.~G. Meier, M. Kronseder, K. Chadova, S. Wimmer, D. Ködderitzsch, H. Ebert \textit{ et al.}, Phys. Rev. Lett. \textbf{117}, 167204 (2016).
    
\bibitem{ywu2019} Y. Xu, Y. Yang, H. Xie, and Y. Wu, Appl. Phys. Lett. \textbf{115}, 182406 (2019).

\bibitem{Qi18} C. Zhou, Y.~P. Liu, Z. Wang, S.~J.~Ma, M.~W. Jia, R.~Q. Wu, L. Zhou, W. Zhang, M.~K. Liu, Y.~Z. Wu, and J. Qi Phys. Rev. Lett. \textbf{121}, 086801 (2018).

\bibitem{MoS2} L. Cheng, X. Wang, and W.~T. Yang, Nat. Phys. \textbf{15}, 347 (2019).

\bibitem{gorchon2017} R.~B. Wilson, Y. Yang, J. Gorchon, C.-H. Lambert, S. Salahuddin, and J. Bokor, Phys. Rev. \textbf{B96}, 045105 (2017).

\bibitem{wu2019} Q. Zhang, Z. Luo, H. Li, Y. Yang, X. Zhang, and Y. Wu, Phys. Rev. Applied \textbf{12}, 054027 (2019).

\bibitem{Cerqueira2019} C. Cerqueira, J.~Y. Qin, H. Dang, A. Djeffal, J.-C. Le Breton, M. Hehn, J.-C. Rojas-Sanchez, X. Devaux, S. Suire, S. Migot \textit{et al.}, Nano Letters \textbf{19}, 90 (2019).
    
\bibitem{iguchi2016} R. Iguchi and E. Saitoh, Journal of the Physical Society of Japan \textbf{86}, 011003 (2017).

\bibitem{bocklage1} L. Bocklage, Sci. Rep. \textbf{6}, 227127 (2016).

\bibitem{bocklage2} L. Bocklage, Phys. Rev. Lett. \textbf{118}, 257202 (2017).

\bibitem{maldonado2017} P. Maldonado, K. Carva, M. Flammer, and P.~M. Oppeneer, Phys. Rev. \textbf{B96}, 174439 (2017).

\bibitem{battiato2010} M. Battiato, K. Carva, and P.~M. Oppeneer, Phys. Rev. Lett. \textbf{105}, 027203 (2010).

\bibitem{battiato_theory_2012} M. Battiato, K. Carva, and P.~M. Oppeneer, Phys. Rev. \textbf{B86}, 024404 (2012).

\bibitem{valetfert1993} T. Valet and A. Fert, Phys. Rev. \textbf{B48}, 7099 (1993).

\bibitem{kaltenborn2012} S. Kaltenborn, Y.-H. Zhu, and H. C. Schneider, Phys. Rev. B 85, 235101 (2012).

\bibitem{nenno2019} D.~M. Nenno, R. Binder, and H. C. Schneider, Physical Review Applied 11, 054083 (2019).

\bibitem{ASA} O. Gunnarsson, O. Jepsen, and O. K. Andersen, Phys. Rev. \textbf{B27}, 7144 (1983).

\bibitem{Turek} I. Turek, V. Drchal, J. Kudrnovsky, M. Sob, and P. Weinberger, \textit{Electronic Structure of Disordered Alloys, Surfaces and Interfaces}, (Springer, New York, 1997).

\bibitem{Questaal} D. Pashov, S. Acharya, W. R. L. Lambrecht, J. Jackson, K. D. Belashchenko, A. Chantis, F. Jamet, and M. van Schilfgaarde, arXiv:1907.06021 (2019).

\bibitem{Barth_1972} U. von Barth and L. Hedin, Journal of Physics C: Solid State Physics \textbf{5}, 1629 (1972).

\bibitem{Schep1997} K.~M. Schep, J. van Hoof, P.~J. Kelly, G. E. W. Bauer, and J.~E. Inglesfield, Phys. Rev. \textbf{B56}, 10805 (1997).

\bibitem{dang2020} T.~H. Dang, Q. Barbedienne, D. Q. To, E. Rongione, N. Reyren, F. Godel, S. Collin, J. M. George, and H. Jaffrès, Phys. Rev. B \textbf{102}, 144405 (2020).    
    
\bibitem{wieczorek_separation_2015} J. Wieczorek, A. Eschenlohr, B. Weidtmann, M. Rösner, N. Bergeard, A. Tarasevitch, T. O. Wehling,
and U. Bovensiepen, Phys. Rev. \textbf{B92}, 174410 (2015).

\bibitem{bass_spin-diffusion_2007} J. Bass and W. P. Pratt, Journal of Physics: Condensed Matter \textbf{19}, 183201 (2007).

\bibitem{gall_electron_2016} D. Gall, J. of Appl. Phys. \textbf{119}, 085101 (2016).

\bibitem{zhukov2006} V.~P. Zhukov, E.~V. Chulkov, and P.~M. Echenique, Phys. Rev. \textbf{B73}, 125105 (2006).

\bibitem {fert_theory_1996} A. Fert and S.-F. Lee, Phys. Rev. \textbf{B53}, 6554 (1996).

\bibitem{hughes2012} J. Lloyd-Hughes and E. Saitoh, Journal of Infrared, Millimeter, and Terahertz Waves \textbf{33}, 871 (2012).






  



\end{thebibliography}
\end{document}